\RequirePackage[l2tabu, orthodox]{nag}
\RequirePackage{snapshot}

\documentclass[9pt,onecolumn]{extarticle}

\makeatletter
\def\@font@info#1{}
\makeatother

\makeatletter
\if@twocolumn
  \usepackage[dvips,letterpaper,top=0.5in, bottom=0.5in, left=0.75in, right=0.5in,includefoot,heightrounded]{geometry}
\else
  \usepackage[dvips,letterpaper,margin=1in,includefoot,heightrounded]{geometry}
\fi

\usepackage{srcltx}

\usepackage[russian,portuges,english]{babel}

\iflanguage{portuges}
    {\newcommand{\keywordname}{Palavras-chaves}}
    {\newcommand{\keywordname}{Keywords}}

\usepackage{amsmath}
\usepackage{amssymb}
\usepackage{amsfonts}
\usepackage{mathtools}
\usepackage{amsthm}

\usepackage{abstract}

\usepackage{graphicx}
\usepackage[usenames,dvipsnames,svgnames,x11names]{xcolor}
\usepackage{subfig}
\usepackage{psfrag}

\usepackage{booktabs}

\usepackage{setspace}
\usepackage{flushend}

\usepackage{multicol}

\usepackage{cite}

\usepackage{hyperref}
\usepackage[normalem]{ulem}
\usepackage{soul}

\usepackage{enumerate}

\usepackage{multirow}
\usepackage[noend]{algpseudocode}

\usepackage{algorithm}

\usepackage{listings}

\usepackage{fancyvrb}

\usepackage[short,12hr]{datetime}

\usepackage{siunitx}
       \usepackage{fouriernc}

\newcommand{\printtitle}{%
\makeatletter
\if@twocolumn

\twocolumn[%
  \maketitle
  \begin{onecolabstract}
    \myabstract
  \end{onecolabstract}
  \begin{center}
    \small
    \textbf{\keywordname}
    \\\medskip
    \mykeywords
  \end{center}
  \bigskip
]
\saythanks
\else
  \maketitle
  \begin{onecolabstract}
    \myabstract
  \begin{center}
    \small
    \textbf{\keywordname}
    \\\medskip
    \mykeywords
  \end{center}
  \end{onecolabstract}
  \bigskip
  \onehalfspacing
\fi
\makeatother
}

\author{%
L.~Portella%
\thanks{L. Portella was with the Department of Statistics, Universidade \mbox{Estadual} de Campinas, Campinas 13083-859, Brazil and the Industrial Signal Processing Laboratory, Universidade Federal de Pernambuco, Caruaru, Brazil (e-mail: \mbox{luanps@unicamp.br).}}%
\and
F.~M.~Bayer%
\thanks{F. M. Bayer is with the Department of Statistics and LACESM, Universidade Federal de Santa Maria, Santa Maria 97105-900, Brazil and the Department of Mathematics and Natural Sciences, Blekinge Institute of Technology, Karlskrona, 37179, Sweden
(e-mail: bayer@ufsm.br).}
\and
R.~J.~Cintra%
\thanks{R. J. Cintra is with
	the Industrial Signal Processing Laboratory,
	Department of Technology,
	Universidade Federal de Pernambuco,
	Caruaru 55014-900, Brazil
  (e-mail: rjdsc@de.ufpe.br).}
}

\title{%
Multiplierless DFT Approximation Based on the Prime~Factor Algorithm}

\newcommand{\myabstract}{%
Matrix approximation methods
have successfully
produced
efficient, low-complexity
approximate transforms
for the
discrete cosine transforms
and
the
discrete Fourier transforms.
For the DFT case,
literature
archives
approximations
operating
at
small power-of-two
blocklenghts, such as \{8, 16, 32\},
or
at large blocklengths,
such as 1024,
which are obtained
by means of the Cooley-Tukey-based approximation
relying on the small-blocklength approximate transforms.
Cooley-Tukey-based approximations
inherit the intermediate multiplications
by twiddled factors
which are usually not approximated;
otherwise
the effected error propagation
would prevent the overall good performance of the approximation.
In this context,
the prime~factor algorithm
can furnish the necessary framework
for
deriving
fully multiplierless DFT approximations.
We introduced
an approximation method
based
on small prime-sized
DFT approximations
which
entirely
eliminates
intermediate multiplication steps
and
prevents
internal error propagation.
To demonstrate the proposed method,
we design
a fully multiplierless
1023-point DFT approximation
based
on 3-, 11- and 31-point DFT approximations.
The performance evaluation
according to popular metrics
showed
that
the proposed approximations
not only
presented
a
significantly lower arithmetic complexity
but
also
resulted
in
smaller approximation error measurements
when
compared to competing methods.
}

\newcommand{\mykeywords}{%
Fast algorithms,
approximate DFT,
multiplicative complexity,
prime factor algorithm.
}

\date{}

\begin{document}

\printtitle

\section{Introduction}

{T}{he} discrete Fourier transform (DFT)
is
a central tool in signal processing~\cite{bracewell1986fourier},
finding applications in a very large number of contexts,
such
as
spectral analysis~\cite{stein2000digital},
filtering~\cite{oppenheim1999discrete},
data compression~\cite{rao2018transform},
and
fast convolution~\cite{brigham1988fast},
to cite a few.
The widespread usage of the DFT
is
due to its rich physical interpretation~\cite{briggs1995dft}
and
the existence of efficient methods
for its computation~\cite{myers1990digital}.
Although
the direct computation of the $N$-point {DFT}
is
an operation in $\mathcal{O}(N^2)$---which
is
prohibitively expensive~\cite{heideman1988multiplicative}---%
efficient algorithms~\cite{burrusdft,briggs1995dft,myers1990digital}
collectively known
as
fast Fourier transforms (FFTs)~\cite{blahut2010fast}
are
capable
of evaluating the DFT
with much less numerical operations
placing the resulting complexity in $\mathcal{O}(N \log N)$ ~\cite{burrusdft}.

Despite
such substantial reduction in complexity,
the remaining operations
can
still be significant
in contexts
where
severe restrictions in computational power
and/or
energy autonomy
are
present~\cite{lee2009energy}.
Such restrictive conditions
arise
in the
framework of wireless communication~\cite{maharatna200464,zhang2015large},
embedded systems~\cite{kitamura2001copyright,8342166},
and Internet of Things (IoT)~\cite{stankovic2014research,lu2014connected}.

Inspired
by the successful methods
for approximating the discrete cosine
transform~\cite{britanak2010discrete,
haweel2001new,
cintra2011integer,
cintra2014low,bayer2012dct,cintra2011dct,
bouguezel2008low,bouguezel2010novel
},
in~\cite{villagran2015aproximaccoes},
a suite of multiplierless DFT approximations
was
derived
for $N=8$, $16$, and $32$~\cite{cintra2024approximation,suarez2014multi,kulasekera2015multi,ariyarathna2018multibeam,madanayake2019towards,madanayake2020fast}.
These {DFT} approximations
were
demonstrated
to provide
spectral estimates
close to
the exact DFT computation,
while
requiring
only
26, 54, and 144 additions
for real-valued input,
respectively~\cite{cintra2024approximation,villagran2015aproximaccoes}.
Broadly,
finding
approximate transforms that closely match the performance of the exact ones
is
a hard task,
because
it
is
often posed
as an integer non-linear matrix optimization problem
with a large number of variables~\cite{ehrgott2005multicriteria}.
Thus,
as $N$ increases,
obtaining
good approximations
becomes
an exceedingly demanding problem to be solved~\cite{portella2022radix}.
As a consequence,
designers of DFT approximations
make use
of indirect methods
such as
(i)~mathematical relationships
between small-sized
and
large-sized DFT matrices~\cite{burrusdft},
(ii)~matrix functional recursions~\cite{nussbaumer1981fast},
and (iii)~matrix decompositions~\cite{seber2008matrix}.
The systematic derivation of good DFT approximations for large block sizes
is
still an open problem
and technical advances
occur
in a case-by-case fashion
due to the inherent numerical difficulties of finding integer matrices
that ensure competitive performance.

Following
such an indirect approach,
the 32-point DFT approximation
discussed in~\cite{cintra2024approximation,villagran2015aproximaccoes}
was
employed
as the fundamental block of the 1024-point DFT approximation
introduced
in~\cite{madanayake2020fast}.
When employed
as a fundamental block
to obtain larger transforms,
the DFT
is referred to as a
ground transformation.
The methodology
described in~\cite{madanayake2020fast}
revisits
the Cooley-Tukey algorithm
and effectively
extends
a given 32-point DFT approximation
resulting
in a $32^2$-point DFT approximation.
This extension
stems
from the fact that the Cooley-Tukey algorithm
can
be formulated
according to a two-dimensional mapping
such
that
the computation of the 1024-point DFT
is
performed
by $2\times 32$ instantiations of the 32-point DFT~\cite{duhamel1990fast}.
However,
even
considering
multiplierless 32-point DFT approximations,
the resulting 1024-point DFT approximations proposed in~\cite{madanayake2020fast}
are
not multiplication-free.
Indeed,
the Cooley-Tukey-based approximations
inherit
the twiddle factors
present
in the exact formulation of the traditional Cooley-Tukey algorithm~\cite{myers1990digital}.
Thus,
the final resulting arithmetic complexity of the best Cooley-Tukey-based 1024-point DFT approximation in~\cite{madanayake2020fast}
is
2883 real multiplications
and 25155 additions,
which
represent
approximately a 72\% reduction in terms of real multiplication
and an 18\% reduction in terms of additions
when
compared to
the exact Cooley-Tukey algorithm~\cite{blahut2010fast}.

The goal of the present paper
is
to propose
a framework
for
deriving
large DFT approximations
that
are
fully multiplierless.
Although,
in fixed-point arithmetic,
any multiplication can theoretically be expressed as a sum of dyadic terms,
we
adopt
the term multiplierless in a more restrictive and practical sense,
consistent with~\cite{britanak2010discrete},
where the minimum number of adders is sought.
In this work,
matrices elements
assumes
values in $\{0,\pm1,\pm\frac{1}{2}\}$
and
eventual scaling constants
have their dyadic representation
limited
to at most
two additions.
Such multiplierlessness criterion
emphasizes
that the proposed approximations rely only on additions and bit-shifting operations,
aiming at the minimum number of adders,
so that future implementations can achieve
reductions in chip area,
power consumption,
and
delay~\cite{ariyarathna2018multibeam}.
For such an end,
we
aim
at exploiting
the prime~factor algorithm (PFA)~\cite{good1958interaction,thomas1963using},
also known
as the Good-Thomas algorithm.
The PFA
has
distinct number-theoretical properties
capable
of
performing the DFT computation
without
intermediate computations
such as
twiddle factors,
which the traditional radix-2 algorithm heavily rely on.
This approach
allows
the construction of scalable, multiplierless DFT approximations,
significantly reducing arithmetic complexity while maintaining competitive approximation accuracy.

Due to the very number-theoretical nature of the PFA,
the resulting transform blocklength
cannot be a power-of-two~\cite{blahut2010fast}.
However,
the design
of non-power-of-two DFT methods~\cite{5495517,van2009non,992857}
is
a promising topic
in
(i)~beamforming and direction of arrival~(DOA)
estimation~\cite{8661061,vaidyanathan2010sparse,qin2015generalized,alawsh2018multi,wang2018unified,shi2017generalized};
(ii)~5G~broadcasting
which
usually present
\mbox{$2^n\cdot3^m$}-point ($n\leq 11$) input signals~\cite{9325758};
(iii)~hybrid algorithms~\cite{9442773,4711206};
(iv)~when the length cannot be chosen, such as in
digital radio technology~~\cite{lai2013high,kim2008design,hofmann2003digital,etsi2009digital};
(v)~channel equalization~\cite{8352390,pfapatent};
and
(vi)~scenarios that presents flexible or adaptive transform lengths as LTE~\cite{chelliah2020power}
and MIMO-OFDM systems~\cite{6195031}.
For examples,
FFT blocklengths like 128, 512, 1024, and 2048,
which are usually implemented with radix-2 algorithms,
could be replaced by alternative lengths such as
130 ($2\times5\times13$),
510 ($2\times3\times5\times17$),
1023 (which is adopted in this work to demonstrate the proposed method),
and 2046 ($2\times3\times11\times31$), which are compatible with the PFA.
Although
power-of-two DFT algorithms
are
more common,
in the end,
the total number of operations
might
play
a decisive role.

As a consequence,
we
separated
the particular blocklegth $N=1023$
as a representative case study to highlight the benefits of the proposed approach
in contrast with the state-of-the-art in large-scale DFT approximations introduced in~\cite{madanayake2020fast}.
The comparison
justified by
the fact that in~\cite{madanayake2020fast}
it is provided a practical implementation benchmark
for low-power applications such as beamforming, reinforcing the relevance of a direct comparison.
Other multiplierless strategies based on twiddle-factor approximations, such as those using sum-of-powers-of-two (SOPOT) coefficients~\cite{chan2002efficient}
and streaming multiplierless FFTs (SMUL-FFT)~\cite{mirfarshbafan2021smul}, are also effective in specific contexts.
However, these methods tend to be more suitable for small-to-medium transform sizes.
As the blocklength increases,
the number of distinct twiddle-factor matrices to be approximated grows rapidly,
shifting the computational burden from multiplications to managing a large number of additive operations or fixed-pattern multipliers.
This can significantly reduce their practical efficiency in large-scale settings.

It
is
worth emphasizing that,
although this work focuses on the 1023-point DFT approximation to enable a direct comparison with the method in~\cite{madanayake2020fast},
the proposed approach
is
built upon the PFA,
which
naturally inherits scalability to any blocklength that can be factorized into coprime factors.
This scalability
stems
from the Chinese Remainder Theorem,
which underlies the Good–Thomas mapping and enables the decomposition of large transforms into smaller ground transforms.
Furthermore,
it
is
also possible to develop hybrid algorithms where radix-2 process the power-of-two factors,
while the proposed PFA-based multiplierless method
computes the remaining coprime factors.

The paper
is
organized
as follows.
Section~\ref{S:mathbackground}
provides
an overview of the DFT and the PFA.
Section~\ref{S:methodology}
describes
the methodology
to obtain
the proposed DFT approximation.
In Section~\ref{S:approximation},
the proposed approximations and algorithm
are
detailed.
In Section~\ref{S:1023app},
the proposed method
is employed
to propose
a multiplierless 1023-point DFT approximation.
Fast algorithms and arithmetic complexity
are
presented in Section~\ref{S:fastalgorithm}.
Section~\ref{companddisc}
reports
the error analysis of the approximations.
In Section~\ref{S:conclusion},
conclusions
are
summarized.

\section{Mathematical background}\label{S:mathbackground}

In this section,
we
review
the DFT mathematical background
and
briefly describe the prime~factor algorithm.

\subsection{Definition of the DFT}

The DFT
is
a linear transformation
that
maps
an $N$-point discrete signal $\mathbf{x}=[x[0],x[1],\ldots, x[N-1]]^ \top$
into
an output signal $\mathbf{X}=[X[0],X[1],\ldots, X[N-1]]^ \top$
by means of
the following expression~\cite{blahut2010fast}:
\begin{align}
	X[k]\triangleq \sum_{n=0}^{N-1}\omega_{N}^{nk}\cdot x[n],\quad k=0,1,\ldots,N-1,
\end{align}
where $X[k]$
is
the $k$th DFT coefficient,
$\omega _{N}^{ }=e^{-j\frac{2 \pi}{N}}$
is
the $N$th root of unity,
and $j\triangleq \sqrt{-1}$.
Although,
the input signal $\mathbf{x}$
may be
real or complex,
this work focuses on the general case of complex-valued inputs.

The DFT
can
also be
expressed
in matrix format according to the next expression:
\begin{align*}
	\mathbf{X}=\mathbf{F}_N \cdot \mathbf{x}
	,
\end{align*}
where $\mathbf{F}_N$
is
the DFT matrix defined by
\begin{align*}
	\mathbf{F}_N=
	\begin{bmatrix}
		1&1&1&\ldots&1\\
		1&\omega_{N}&\omega^{2}_{N}&\ldots&\omega^{N-1}_{N}\\
		1&\omega^{2}_{N}&\omega^{4}_{N}&\ldots &\omega^{2(N-1)}_{N}\\
		1&\omega^{3}_{N}&\omega^{6}_{N}&\ldots &\omega^{3(N-1)}_{N}\\
		\vdots&\vdots&\vdots&\ddots&\vdots\\
		1&\omega^{N-1}_{N}&\omega^{2(N-1)}_{N}&\ldots&\omega^{(N-1)(N-1)}_{N}\\
	\end{bmatrix}
	.
\end{align*}

\subsection{Prime~Factor Algorithm}\label{S:PFA}

Comparable to
the more popular Cooley-Tukey FFT \cite{blahut2010fast},
the PFA
is
a factorization-based FFT
capable of
computing the $N$-point DFT,
where $N=N_1 \times N_2$,
with $N_1$ and $N_2$ being relatively prime,
i.e., $\operatorname{gcd}(N_1,N_2)=1$.
The method
is
based on
a number-theoretical re-indexing~\cite{myers1990digital} of the input signal coefficients
into a two-dimensional array~\cite{oppenheim1999discrete}
which
is
based on
the Chinese remainder theorem~\cite{blahut2010fast}.

The PFA
is
summarized
according to
the following description\cite[Fig.~3.8]{blahut2010fast}:
\begin{enumerate}[(i)]
	\item Obtain $n_1$ and $n_2$
	that
	satisfy $ (n_1\cdot N_1+ n_2\cdot N_2) \mod N=1$ \cite{myers1990digital};

	\item \label{map} Map $\mathbf{x}$
	into a block of size $N_1\times N_2$
	according to
	the following 1D to 2D rearrangement of elements:
	\begin{align*}
			\operatorname{map}\left(
\begin{bmatrix}
x{[0]}\\
x{[1]}\\
x{[2]}\\
\vdots\\
x{[N-1]}\\
\end{bmatrix}
\right)
=
\begin{bmatrix}
x{[0]}&x{[r] }&\ldots&x{[(N_2 - 1)r]}\\
x{[s]}&x{[r + s] }&\ldots&x{[r + (N_2 - 1)r] }\\
\vdots&\vdots&\ddots&\vdots\\
x{[(N_1 - 1)s]}&x{[(N_1 - 1)r + s]}&\cdots&x{[(N_1 - 1)s + (N_2 - 1)r]}\\
\end{bmatrix}
			,
	\end{align*}
	where $r=N_1\cdot n_1$ and $s=N_2\cdot n_2$;
	\item \label{column}Compute
	the $N_2$-point DFT of each column of the 2D array
	obtained
	in Step 2);
	\item \label{rown}Compute
	the $N_1$-point DFT of each row of the resulting 2D array from Step 3);
	\item \label{invmap}Reconstruct
	the vector $\mathbf{X}$
	from the resulting block according to the following mapping:
	\begin{align*}
			\operatorname{invmap}\left(
\begin{bmatrix}
					X{[0]}&X{[N_1]}& \cdots &X{[(N_2 - 1)N_1]}\\
					X{[N_2]}&X_{[N_1 + N_2] }& \cdots &X{[N_2 + (N_2 - 1)N_1]}\\
					\vdots&\vdots&\ddots&\vdots\\
					X{[(N_1 - 1)N_2]}&X{[(N_1 - 1)N_2 + N_1]}& \cdots &X{[(N_1 - 1)N_2 + (N_2 - 1)N_1]}\\
\end{bmatrix}
\right)
			=
\begin{bmatrix}
				X{[0]}\\
				X{[1]}\\
				X{[2]}\\
				\vdots\\
				X{[N - 1]}\\
\end{bmatrix}
			.
	\end{align*}

\end{enumerate}

All index operations
are
performed in modulo $N$ arithmetic,
ensuring
the correct size of the arrays.
The algorithm
can
be synthesized
as follows:
\begin{align}\label{PFA}
	\mathbf{X}=\mathrm{invmap}\left( \mathbf{F}_{N_1}\cdot \left[\mathbf{F}_{N_2} \cdot(\mathrm{map}(\mathbf{x}))^\top\right]^\top\right)
	.
\end{align}

Notice that
if $N_1$ or $N_2$
can
be factored
into relative primes,
then
the algorithm
can
be reapplied.
The $N_1$- and $N_2$-point transformations
are
referred to
as ground transformations.

\section{Approximate DFT methodology }\label{S:methodology}

Being
alternatives
to the exact transformations,
approximate transforms
possess
a low computational cost
and provide similar mathematical properties
and performance
to their exact counterparts.
In this context,
an approximate DFT matrix ${\hat{\mathbf{F}}}_N^{*}$
can
be derived
by solving the following optimization problem:
\begin{align}
	\label{general}
	\hat{\mathbf{F}}_N^{*}=\underset{\hat{\mathbf{F}}_N}{\arg\min\operatorname{error}} (\hat{\mathbf{F}}_N,\mathbf{F}_N)
	,
\end{align}
where $\operatorname{error}(\cdot)$
represents
the adopted error measure
and $\hat{\mathbf{F}}_N$
is
a candidate approximation
obtained
from a suitable search space.

Approximate transforms
can
be derived
from low-complexity matrices~\cite{cintra2011integer}
according to
an orthogonalization process
referred to
as polar decomposition~\cite{higham1986computing}.
Such approach consists of two matrices:
a low-complexity matrix and a real-valued diagonal matrix.
It
is
important to note that
an auxiliary method
is
required to generate the low-complexity matrix
because
the polar decomposition
alone does not provide
such a matrix.
If
the polar decomposition
is applied directly
to the exact DFT matrix or to an already orthogonal approximation,
the resulting diagonal matrix
becomes
the identity matrix.
The auxiliary method
to obtain the low-complexity matrix
is
presented in the next section.

A candidate approximation $\hat{\mathbf{F}}_N$
for the exact transformation $\mathbf{{F}}_N$
can
be written
as:
\begin{align}
	\hat{\mathbf{F}}_N=\sqrt{N}\cdot\mathbf{S}_N \cdot {\mathbf{T}}_N
	,
\end{align}
where $\mathbf{T}_N$
is
a low-complexity matrix
and $\mathbf{S}_N$
is
a diagonal matrix expressed by
\begin{align}
	\label{S}
	\mathbf{S}_N=\operatorname{diag}\left(\sqrt{\left[\operatorname{diag}\left({\mathbf{T}}_N \cdot {\mathbf{T}}_N^{H}\right)\right]^{-1}}\right)
	,
\end{align}
being $\operatorname{diag}(\cdot)$
a function
that
returns
a diagonal matrix,
if the argument
is
a vector;
or
a vector with the diagonal elements,
if the argument
is
a matrix,
the superscript ${}^H$
denotes
the Hermitian operation~\cite{seber2008matrix},
and $\sqrt{\cdot}$
is
the matrix square root operation~\cite{higham1987computing}.
Therefore,
a suitable choice of $\mathbf{T}_N$
is
central to the above approach.
Thus,
from this point onward,
we focus on the derivation of $\mathbf{T}_N$.
For simplicity of notation,
the constant $\sqrt{N}$
is
absorbed into the matrix $\mathbf{S}_N$
as follows
\begin{align}
	\label{decomp}
	\hat{\mathbf{F}}_N=\hat{\mathbf{S}}_N \cdot {\mathbf{T}}_N
	,
\end{align}
where $\hat{\mathbf{S}}_N =\sqrt{N}\cdot\mathbf{S}_N$.

\subsection{Search Space}\label{search-space}

The low-complexity matrices $\mathbf{T}_N$
are
taken
from the search space
given by
the matrix space $\mathcal{M}_N (\mathcal{P})$,
which
is
the set of all $N \times N$ matrices
with entries over a set of low-complexity multipliers $\mathcal{P}$.
Popular choices for $\mathcal{P}$
are
$\lbrace-1,0,1\rbrace$
and $\lbrace-1,-\frac{1}{2},0,\frac{1}{2},1\rbrace$,
which
contain
only trivial multipliers~\cite{blahut2010fast}.

The set $\mathcal{M}_N (\mathcal{P})$
can
be
extremely large.
For instance,
$\mathcal{M}_N (\mathcal{P})$
contains
$3^{64}\approx3.43\times 10^{30}$ elements~(distinct matrices) for $N=8$
and $\mathcal{P}=\lbrace-1,0,1\rbrace$.
Therefore,
we
propose
as a working search space,
a subset of $\mathcal{M}_N (\mathcal{P})$
given by
the expansion factor methodology~\cite{britanak2010discrete}.
Thus,
low-complexity matrices $\mathbf{T}_N$
can
be generated
according to the following expression:
\begin{align}
	\label{expansion}
	\mathbf{T}_N=g(  \alpha \cdot \mathbf{F}_N)
	,
\end{align}
where $g(\cdot)$
is
an entry-wise integer matrix function,
such as rounding, truncation, ceiling, and floor functions~\cite{cintra2011integer}
and
$\alpha$
is
a real number
referred to
as the expansion factor~\cite{malvar2002low}.
To ensure
that the integer function $g(\cdot)$
returns
only values
within
$\mathcal{P}$,
the values of $\alpha$
are
judiciously restricted
to an interval $\mathcal{D}$ given by
\begin{align}
	\label{alphainterval}
	\alpha_\text{min}\leq\alpha\leq\alpha_\text{max},
\end{align}
where
$\alpha_\text{min}=\inf\lbrace\alpha \in \mathbb{R}_{+}: g(\alpha\cdot \gamma_\text{max})\neq0\rbrace$
and $\alpha_\text{max}=\sup\lbrace\alpha \in \mathbb{R}_{+}: g(\alpha\cdot \gamma_\text{max})= \max(\mathcal{P})\rbrace$,
being $\gamma_\text{max}=\underset{m,n}{\max}(|\Re(f_{m,n})|,|\Im(f_{m,n})|)$
and $f_{m,n}$, the $(m,n)$th entry of $\mathbf{F}_N$.
The symmetries of $\mathbf{{F}}_N$
allow
us
to restrict
the analysis to $\alpha \geq 0$
and since the entries of $\mathbf{{F}}_N$
are
bounded
by the unity,
$\gamma_\text{max}=1$.

\subsection{Optimization Problem and Objective Function}\label{S:optimPro}

The general optimization problem
shown
in~\eqref{general}
can be formulated as
\begin{align}
	\alpha^*=&\underset{\alpha \in \mathcal{D}}{\arg\min\operatorname{error}}(\hat{\mathbf{F}}_N,\mathbf{F}_N)
	,
\end{align}
employing~\eqref{decomp}
\begin{align}
	\alpha^*=&\underset{\alpha \in \mathcal{D}}{\arg\min\operatorname{error}}(\hat{\mathbf{S}}_N\cdot {\mathbf{T}}_N,\mathbf{F}_N)
	,
\end{align}
and
rewritten
applying~\eqref{expansion}
\begin{align}
	\label{optm}
	\alpha^*=&\underset{\alpha \in \mathcal{D}}{\arg\min\operatorname{error}}(\hat{\mathbf{S}}_N\cdot g(  \alpha \cdot \mathbf{F}_N),\mathbf{F}_N)
	.
\end{align}

Therefore,
the low-complexity matrix
is
obtained by
\begin{align}
{\mathbf{T}}_N^*=g(  \alpha^* \cdot \mathbf{F}_N)
,
\end{align}
and the optimal approximation
is
furnished by
\begin{align}
	\label{optF}
	\hat{\mathbf{F}}_N^*=\hat{\mathbf{S}}_N^*\cdot {\mathbf{T}}_N^*
	,
\end{align}
where
$\hat{\mathbf{S}}_N^*$
stems
from ${\mathbf{T}}_N^*$
as detailed in~\eqref{S},
\textit{mutatis mutandis}.

Now
we
aim at
specifying the error function in~\eqref{optm}.
As shown
in literature~\cite{potluri2012multiplier,madanayake2019towards,tablada2015class},
usual choices for such function
are:
(i)~the total error energy~\cite{cintra2011dct};
(ii)~the mean absolute percentage error~(MAPE)~\cite{everitt2002cambridge};
and (iii)~the deviation from orthogonality~\cite{cintra2014low,flury1986algorithm}.
These metrics
are
described below.
\begin{enumerate}[(i)]
	\item The total error energy
	is
	defined by
	\begin{align*}
		\epsilon({\hat{\mathbf{F}}_N})=\pi\cdot||\mathbf{F}_N-\hat{\mathbf{F}}_N||_{\text{F}}^{2}
		,
	\end{align*}
	where
	$||\cdot||_{\text{F}}$
	represents
	the Frobenius norm~\cite{watkins2004fundamentals};
	\item The MAPE of the transformation matrix
	is
	obtained by
	\begin{align*}
		M({\hat{\mathbf{F}}_N})=100\cdot\frac{1}{N^2}\cdot\sum_{m=1}^{N}\sum_{n=1}^{N}\left|\frac{f_{m,n}-\hat{f}_{m,n}}{f_{m,n}}\right|
		,
	\end{align*}
	where
	$\hat{f}_{m,n}$
	is
	the $(m,n)$th entry of $\hat{\mathbf{F}}_N$;
	\item The deviation from orthogonality~\cite{cintra2014low}
	is
	defined by:
	\begin{align*}
		\phi(\hat{\mathbf{F}}_N)=1-\frac{||\operatorname{diag}(\mathbf{\hat{\mathbf{F}}}_N\cdot\hat{\mathbf{F}}_N^{H})||_{\text{F}}}{||\mathbf{\hat{\mathbf{F}}}_N\cdot\hat{\mathbf{F}}_N^{H}||_{\text{F}}}
		.
	\end{align*}

	Small values of $\phi(\cdot)$
	indicate
	proximity to orthogonality.
	Orthogonal matrices
	have
	null deviation.
\end{enumerate}

Combining
the above error functions
in a single optimization problem,
we
obtain
the following multicriteria problem~\cite{ehrgott2005multicriteria,barichard2008multiobjective}:
\begin{align}
		\alpha^*=\underset{\alpha\in\mathcal{D}}
		{\arg\min}
		\Bigl\{\epsilon\Big(\hat{\mathbf{S}}_N^*\cdot g(\alpha\cdot \mathbf{F}_N)\Big),
		M\Big(\hat{\mathbf{S}}_N^*\cdot g(  \alpha \cdot \mathbf{F}_N)\Big),
		\phi\Big(\hat{\mathbf{S}}_N^*\cdot g(\alpha\cdot\mathbf{F}_N)\Big)\Bigr\}
		.
\end{align}

\section{Multiplierless Prime~Factor Approximation}\label{S:approximation}

In this section,
we
formalize
the mathematical structure
of
the approximate DFT
based
on the
prime~factor algorithm.
Additional to the main structure,
we describe
two variations of the method:
(i)~the unscaled  approximation
and
(ii)~the hybrid approximation.

\subsection{Mathematical Definition}

Under the assumption of the PFA,
we
compute
the $N$-point DFT approximation
as follows
\begin{align}
	\label{apfa}
	\hat{\mathbf{X}}=\mathrm{invmap}\left(\hat{\mathbf{F}}^{*}_{N_1}\cdot\left[\hat{\mathbf{F}}^{*}_{N_2} \cdot(\mathrm{map}(\mathbf{x}))^\top\right]^\top\right)
	,
\end{align}
where
$\hat{\mathbf{F}}^{*}_{N_1}$ and $\hat{\mathbf{F}}^{*}_{N_2}$
are
approximations of ${\mathbf{F}}_{N_1}$
and ${\mathbf{F}}_{N_2}$,
respectively~(cf.~(\ref{PFA})).

If
the approximations used in~\eqref{apfa}
admit
the format expressed in~\eqref{optF},
then
the above equation
can
be rewritten
as
\begin{align}
	\label{APFA}
	\hat{\mathbf{X}}=\mathrm{invmap}\left( \hat{\mathbf{S}}^*_{N_1}\cdot \mathbf{T}^*_{N_1}\cdot \left[\hat{\mathbf{S}}^*_{N_2}\cdot \mathbf{T}^*_{N_2} \cdot(\mathrm{map}(\mathbf{x}))^\top\right]^\top\right)
	.
\end{align}

Notice that
$\hat{\mathbf{{S}}}^*_{N_1}$ and $\hat{\mathbf{{S}}}^*_{N_2}$
are
real diagonal matrices
that
can
be factored out
from the mapping operator
as follows:
\begin{align}
	\label{extract}
	\hat{\mathbf{X}}=\hat{\mathbf{S}}\cdot\mathrm{invmap}\left(\mathbf{T}^*_{N_1}\cdot\left[\mathbf{T}^*_{N_2}\cdot(\mathrm{map}(\mathbf{x}))^\top\right]^\top\right)
	,
\end{align}
where
$\hat{\mathbf{S}}$
is
a diagonal matrix
given by
\mbox{$\hat{\mathbf{S}}=\operatorname{diag}\big[\mathrm{invmap}\left(\operatorname{diag}(\hat{\mathbf{{S}}}^*_{N_1})\cdot \operatorname{diag}(\hat{\mathbf{{S}}}^*_{N_2})^\top\right)\big]$}
.

\subsection{Unscaled Approximation}

Because the matrix $\hat{\mathbf{S}}$
is
diagonal,
its
role in the approximate DFT computation consisits of scaling each spectral component.
Depending on
the context in which the DFT
is
applied,
the scaling
can
be embedded, absorbed, parallel computed, or even neglected
when
the unscaled spectrum
is
sufficient~\cite{johnson2006modified,qadeer2014radix,frigo1997fastest}.
The unscaled $N$-point DFT approximation
is
obtained
by
\begin{align}
	\label{unscaled}
	\tilde{\mathbf{X}}=\mathrm{invmap}\left(\mathbf{T}^*_{N_1}\cdot\left[\mathbf{T}^*_{N_2}\cdot(\mathrm{map}(\mathbf{x}))^\top\right]^\top\right)
	.
\end{align}

Thus,
we
have
the following relationship between~\eqref{extract} and~\eqref{unscaled}:
\begin{align}
	\hat{\mathbf{X}}=\hat{\mathbf{S}}\cdot\tilde{\mathbf{X}}
	.
\end{align}

\subsection{Hybrid Approximations}
\label{S:hybrids}

It might be advantageous to approximate only part of the DFT computation instead of the entire transform.
This approach called hybrid algorithms
allows
for a balance between computational efficiency and accuracy, targeting specific components of the computation for approximation.
In this context,
we
also provide
two hybrid algorithms
approximating
only part of the DFT computation.

First,
we
keep
the row-wise $N_1$-point DFT exact
while
the column-wise $N_2$-point DFT
is
approximated.
The diagonal matrix $\hat{\mathbf{S}}$
can
also be factored out
in the hybrid algorithms.
Thus, this algorithm
is
given by
\begin{align}
	\hat{\mathbf{X}}_1=\hat{\mathbf{S}} \cdot \mathrm{invmap}\left( \left\{\mathbf{F}_{N_1}\cdot \left[\mathbf{T}^*_{N_2} \cdot(\mathrm{map}(\mathbf{x}))\right]^\top\right\}^\top\right)
	,
\end{align}
where
$\hat{\mathbf{S}}=\operatorname{diag}\big[\mathrm{invmap}\left(\mathbf{{1}}_{N_1}\cdot\operatorname{diag}(\hat{\mathbf{{S}}}^*_{N_2})^\top\right)\big]$
and $\mathbf{{1}}_{r}$
is
a column vector of ones
with length
equals
to $r$.

Second,
the column-wise $N_2$-point DFT
is
maintained
exact
and the row-wise $N_1$-point DFT
is
approximated.
Then,
the $N$-point DFT approximation
is
calculated by
\begin{equation}
	\hat{\mathbf{X}}_2=\hat{\mathbf{S}}\cdot\mathrm{invmap}\left(\left\{\mathbf{T}^*_{N_1}\cdot\left[\mathbf{F}_{N_2}\cdot(\mathrm{map}(\mathbf{x}))\right]^\top\right\}^\top\right)
	,
\end{equation}
where
$\hat{\mathbf{S}}=\operatorname{diag}\big[\mathrm{invmap}\left(\operatorname{diag}(\hat{\mathbf{{S}}}^*_{N_1})\cdot (\mathbf{{1}}_{N_2})^\top\right)\big]$.

\section{Approximations for the 1023-point {DFT}}\label{S:1023app}

In this section,
we
advance
two results.
First,
we
apply
the prime~factor algorithm detailed in Section~\ref{S:approximation}
to obtain
approximations for the 1023-point DFT.
Second,
we
apply
the methodology described in Section~\ref{S:methodology}
to obtain
approximations for the 3-, 11- and, 31-point DFTs,
which
are
required
for 1023-point DFT approximations.

\subsection{1023-point DFT Approximation}

Invoking~\eqref{apfa}
for $N_1=31$ and $N_2=33$,
we
introduce
a 1023-point DFT approximation
according to
the following equation:
\begin{align}
	\label{1023apply}
	\hat{\mathbf{X}}=\operatorname{invmap}\left(\hat{\mathbf{F}}^*_{31}\cdot\left[\hat{\mathbf{F}}^*_{33}\cdot(\operatorname{map}(\mathbf{x}))^\top\right]^\top\right)
	.
\end{align}
The term in square brackets in \eqref{1023apply}
requires
31 calls of a 33-point DFT approximation.
Because
$N_2=33=11\times3$
is
suitable
for the proposed method formalism,
a 33-point DFT approximation
can
be obtained
based on approximations for the 3- and 11-point DFTs,
as follows
\begin{align}
	\hat{\mathbf{Y}}=\operatorname{invmap}\left(\hat{\mathbf{F}}^*_{11}\cdot\left[\hat{\mathbf{F}}^*_{3}\cdot(\operatorname{map}(\mathbf{y}))^\top\right]^\top\right)
	,
\end{align}
where
$\mathbf{y}$
is
the 33-point column vector
corresponding to the rows of $\operatorname{map}(\mathbf{x})$.
As shown in~\eqref{extract},
the scaling matrix $\hat{\mathbf{S}}$
can
be calculated separately,
allowing the 1023-point DFT approximation
to be rewritten
as
\begin{align}
	\label{1023dft}
	\hat{\mathbf{X}}=\hat{\mathbf{S}}\cdot\operatorname{invmap}\left(\hat{\mathbf{T}}^*_{31}\cdot\left[\hat{\mathbf{T}}^*_{33}\cdot(\operatorname{map}(\mathbf{x}))^\top\right]^\top\right)
	,
\end{align}
and
\begin{align}
	\label{1023dft-2}
	\hat{\mathbf{Y}}=\operatorname{invmap}\left(\hat{\mathbf{T}}^*_{11}\cdot\left[\hat{\mathbf{T}}^*_{3}\cdot(\operatorname{map}(\mathbf{y}))^\top\right]^\top\right)
	.
\end{align}

Notice
that the diagonal matrix $\hat{\mathbf{S}}$
encompasses
the intermediate diagonals $\hat{\mathbf{{S}}}^*_{3}$, $\hat{\mathbf{{S}}}^*_{11}$, and $\hat{\mathbf{{S}}}^*_{31}$
and
is
given by
\begin{align}
	\hat{\mathbf{S}} \nonumber= \operatorname{diag} \Bigl\{
	\operatorname{invmap} \Big[ \operatorname{diag}(\hat{\mathbf{{S}}}^*_{31})
	\cdot\operatorname{invmap} \Big( \operatorname{diag}(\hat{\mathbf{{S}}}^*_{11}) \cdot \operatorname{diag}(\hat{\mathbf{{S}}}^*_{3})^\top \Big)^\top
	\Big] \Bigr\}
	,
\end{align}
where
$\hat{\mathbf{{S}}}^*_{3}$, $\hat{\mathbf{{S}}}^*_{11}$, and $\hat{\mathbf{{S}}}^*_{31}$
are
the diagonal matrices
required
by the DFT approximations $\hat{\mathbf{F}}^*_{3}$, $\hat{\mathbf{F}}^*_{11}$, and $\hat{\mathbf{F}}^*_{31}$,
respectively,
as described in~\eqref{optF}.

\subsection{Design Parameters}

The algorithm detailed in the previous section
requires
approximations to the 3-, 11-, and 31-point DFT.
To obtain
such approximations,
we
numerically
apply
the methodology described in Section~\ref{S:methodology}
for which
$g(\cdot)$ and $\mathcal{P}$
must
be specified.
As suggested
in~\cite{tablada2015class},
we
define
$\mathcal{P}=\lbrace-1,-\frac{1}{2},0,\frac{1}{2},1\rbrace$ as the set of low-complexity multipliers.
Among the integer functions mentioned,
the $\operatorname{round}$ function,
as implemented in Matlab/Octave~\cite{MATLAB,eaton:octave:2008},
has
been
reported
to offer
superior performance
when compared with other integer functions~\cite{cintra2014low,tablada2015class,cintra2011dct}.
Thus,
as suggested in~\cite{oldham2010atlas},
we
adopted
the following round-to-multiple function:
\begin{align}
	g(x)=\frac{1}{2}\cdot\operatorname{round}(2\cdot x)\in\mathcal{P}
	.
\end{align}

The related $\alpha$ search space
is
$\mathcal{D}=[0.26,1.25]$~(cf.~\eqref{alphainterval}).
The $\alpha$ step used
was
$10^{-5}$
providing
a total of 6, 16, and 42 different approximations
for the 3-, 11-, and 31-point DFTs,
respectively.
Smaller $\alpha$ steps
do not
alter the results.

\subsection{3-, 11-, and 31-point DFT Approximations}

The obtained optimal expansion factors $\alpha^*$
are
in the intervals
$[0.86603,1.25000]$, $[0.99240,1.14528]$, and $[1.08859,1.15141]$
for the 3-, 11-, and 31-point DFT approximations,
respectively.
Therefore,
we
selected
$\alpha^*=\frac{9}{8}$ for convenience.
Then,
the low-complexity matrices
$\mathbf{{T}}^*_{3}$, $\mathbf{{T}}^*_{11}$, and $\mathbf{{T}}^*_{31}$
are obtained by
\begin{align*}
	\mathbf{{T}}^*_{N}=\frac{1}{2}\cdot\operatorname{round}\left(2\cdot\frac{9}{8}\cdot\mathbf{{F}}_{N}\right), \quad N={3,11,31}
.
\end{align*}
From~\eqref{S},
we
obtain
the scaling matrices \( \hat{\mathbf{S}}_N^* \)
according to the following general structure
\begin{align*}
	\hat{\mathbf{S}}_N^* =
	\operatorname{diag}\left(1, \sqrt{\eta_N} \cdot \mathbf{I}_{N-1} \right), \quad N={3,11,31}
	,
\end{align*}
where
$\mathbf{I}_m$
is
the identity matrix of order $m$,
and
the constants
are
$\eta_3 = \frac{6}{7}$,
$\eta_{11} = \frac{11}{13}$,
and $\eta_{31} = \frac{31}{38}$.

Thus,
the 3-, 11- and 31-point DFT approximations
are
obtained by
\begin{align*}
\mathbf{\hat{F}}^*_{N}=\hat{\mathbf{{S}}}^*_{N}\cdot{\mathbf{T}}^*_{N},\quad N={3,11,31}
.
\end{align*}

\subsection{Approximate Scale Factors}

Despite the dramatic reduction in arithmetic complexity,
including the absence of twiddle factors,
the computation shown in~\eqref{1023dft}
requires
\mbox{$2(N-1)$} real multiplications
due to the elements of the diagonal $\hat{\mathbf{S}}$.
The elements of $\hat{\mathbf{S}}$,
$s_i,i=0,1,\ldots,1022$,
are
given by
\begin{align*}
	{s}_{i} = \begin{cases}
		1,&\textrm{if }i = 0,\\
		\sqrt{\frac{6}{7}},&\textrm{if } i\mod 31 = 0\   \wedge  \ i\mod 11 = 0\  \wedge \ i\mod 3 \neq 0 ,\\
		\sqrt{\frac{11}{13}},&\textrm{if } i\mod 31 = 0\  \wedge  \ i\mod 11 \neq 0\  \wedge  \ i\mod 3 = 0,\\
		\sqrt{\frac{66}{91}},&\textrm{if } i\mod 31 = 0\  \wedge  \ i\mod 11 \neq 0\  \wedge  \ i\mod 3 \neq 0,\\
		\sqrt{\frac{31}{38}},&\textrm{if } i\mod 31 \neq 0 \  \wedge  \ i  \mod 11 = 0\  \wedge  \ i\mod 3 = 0,\\
		\sqrt{\frac{93}{133}},&\textrm{if } i\mod 31 \neq 0 \  \wedge  \ i  \mod 11 = 0\  \wedge  \ i\mod 3 \neq 0,\\
		\sqrt{\frac{341}{494}},&\textrm{if } i\mod 31 \neq 0\  \wedge  \ i\mod 11 \neq 0\  \wedge  \ i\mod 3 = 0,\\
		\sqrt{\frac{1023}{1729}},&\textrm{otherwise}.
	\end{cases}
\end{align*}

Further complexity reductions
can be
achieved by
approximating the elements of $\hat{\mathbf{S}}$
using
truncated representations from the canonical signed digit~(CSD) number system~\cite{886740,cintra2025note},
which satisfy the minimum adder representation criterion~\cite{hartley1996subexpression}.
We
approximated
each element in $\hat{\mathbf{S}}$
in such way that
its truncated CSD representation
admits
at most
two additions~\cite{5219227}.
Table~\ref{CSD}
provides
a representation of the elements from $\hat{\mathbf{S}}$,
the absolute error between the approximation, the constants,
and their truncated representation in CSD~($\bar{1}$ represents $-1$).
Since
the elements from $\hat{\mathbf{S}}_{3}$, $\hat{\mathbf{S}}_{11}$, and $\hat{\mathbf{S}}_{31}$
are
present
in $\hat{\mathbf{S}}$,
multiplierless approximation for the 3-, 11-, and 31-point DFTs
are
also possible using Table~\ref{CSD}.
The approximations in which the diagonal matrices
were
also approximated
follow the notation $\hat{\mathbf{F}}^\prime_N$.

\begin{table}
	\caption{Truncated CSD Approximations for the Constants from $\mathbf{S}$}
	\centering
	\begin{tabular}{lccc}
		\toprule
		Constant&Approximation&$|$Error$|$&CSD\\\midrule
		$\sqrt{\frac{66}{91}}\approx0.85163$&$\frac{55}{64}=0.859375$&0.00774&$1.00\bar{1}00\bar{1}0$\\
		$\sqrt{\frac{11}{13}}\approx0.91987$&$\frac{59}{64}=0.921875$&0.00201&$1.000\bar{1}0\bar{1}0$\\
		$\sqrt{\frac{6}{7}}\approx0.92582$&$\frac{119}{128}=0.9296875$&0.00387&$1.000\bar{1}00\bar{1}$\\
		$\sqrt{\frac{341}{494}}\approx0.83083$&$\frac{27}{32}=0.84375$&0.01292&$1.00\bar{1}0\bar{1}00$\\
		$\sqrt{\frac{93}{133}}\approx0.83621$&$\frac{27}{32}=0.84375$&0.00754&$1.00\bar{1}0\bar{1}00$\\
		$\sqrt{\frac{31}{38}}\approx0.90321$&$\frac{29}{32}=0.90625$&0.00304&$1.00\bar{1}0100$\\
		$\sqrt{\frac{1023}{1729}}\approx0.76920$&$\frac{49}{64}=0.765625$&0.00358&$1.0\bar{1}00010$\\
		\bottomrule
	\end{tabular}
	\label{CSD}
\end{table}

\section{Fast Algorithms and Arithmetic Complexity}\label{S:fastalgorithm}

In this section,
we
introduce
fast algorithms
based on sparse matrix factorizations~\cite{blahut2010fast} for
the proposed 3-, 11-, and 31-point DFT approximations
to reduce
the number of remaining operations.
These approximations
are employed
within the PFA to approximately compute the 1023-point DFT.
Although the PFA
has the inherent property of reducing
the total number of operations by nearly half
when the input is real-valued, as shown in~\cite{heideman1984prime},
for the sake of generality and consistency,
the analyses presented in this section
are based on
complex-valued input.
A complex multiplication
can
be translated into
three real multiplications and three real additions~\cite[p.~3]{blahut2010fast}.
The matrices
$\mathbf{T}_N^\ast$,
$N=3,11,31$,
employed in the 1023-point DFT approximations
do not require multiplications;
only additions and bit-shifting operations
are
needed~\cite[p.~221]{britanak2010discrete}.

In the following,
the butterfly-structure matrix
is defined as follows
\begin{align*}
	\mathbf{B}_{m}=
	\left[
	\begin{array}{cc}
		{\mathbf{I}}_{m/2}&\bar{\mathbf{I}}_{m/2}\\
		-\bar{\mathbf{I}}_{m/2}&{\mathbf{I}}_{m/2}
	\end{array}
	\right],
\end{align*}
where
$m$
is
an even integer
and $\bar{\mathbf{I}}_{m/2}$
is
the counter-identity matrix~\cite{horn2012matrix} of order ${m/2}$.

\subsection{3-point Approximation Fast algorithm}\label{3fact}

The low-complexity matrix $\mathbf{T}^*_{3}$
can be
represented as:

\begin{align}
	\label{fast3}
	{\mathbf{T}}^*_{3}=\mathbf{A}_1^\top\cdot\mathbf{C}_1\cdot\mathbf{A}_1
	,
\end{align}
where
\begin{align}
	\mathbf{A}_1 = \operatorname{diag}(1, \mathbf{B}_2)
	,
\end{align}
and
the matrix $\mathbf{C}_1$
is
\begin{align*}
	{\mathbf{C}}_{1} =
	\begin{bsmallmatrix*}[r]\\
		1&1&\\
		1&-\frac{1}{2}&\\
		&&-j\\
	\end{bsmallmatrix*}
	.
\end{align*}
The matrix ${\mathbf{A}}_{1}$
requires
only 4 real additions
and $\mathbf{C}_1$
needs
4 real additions and 2 bit-shifting operations.

Considering
a complex input,
the direct implementation of the $\mathbf{{T}}^*_{3}$
requires
20 additions and 8 bit-shifting operations.
However,
if the factorization in~\eqref{fast3}
is
applied,
then
${\mathbf{T}}^*_{3}$
requires
12 real additions and 2 bit-shifting operations.
To scale
the approximation maintaining the exact $\hat{\mathbf{S}}^*_{3}$,
4 real multiplications
are
added to
the previous arithmetic cost.
This approximation
is
denoted by ${\hat{\mathbf{F}}}^*_{3}$.
Otherwise,
if the diagonal matrix
is
approximated,
8 additions and 8 bit-shifting operations
are
needed
instead of
the multiplications.
This approximation
is
called ${\hat{\mathbf{F}}^{\prime}_{3}}$.

\subsection{11-point Approximation Fast Algorithm}\label{11fact}

For the 11-point approximation,
the low-complexity matrix $\mathbf{T}^*_{11}$
is
given by
\begin{align}
	\label{fast11}
	{\mathbf{T}}^*_{11}=\mathbf{A}_2^\top\cdot\mathbf{C}_2\cdot\mathbf{A}_2
	,
\end{align}
where
\begin{align}
	\mathbf{A}_2 = \operatorname{diag}(1, \mathbf{B}_{10})
	,
\end{align}
and
the block matrix $\mathbf{C}_2$
is
\begin{align*}
\mathbf{C}_2 =
\operatorname{diag}
\left(
\begin{bsmallmatrix*}[r]
	1 & 1 & 1 & 1 & 1 & 1 \\
	1 & 1 & \tfrac{1}{2} &  & -\tfrac{1}{2} & -1 \\
	1 & \tfrac{1}{2} & -\tfrac{1}{2} & -1 &  & 1 \\
	1 &  & -1 & \tfrac{1}{2} & 1 & -\tfrac{1}{2} \\
	1 & -\tfrac{1}{2} &  & 1 & -1 & \tfrac{1}{2} \\
	1 & -1 & 1 & -\tfrac{1}{2} & \tfrac{1}{2} &
\end{bsmallmatrix*}
,
\begin{bsmallmatrix*}[r]
	- j &  j & - j &  \tfrac{j}{2} & -\tfrac{j}{2} \\
	j & -\tfrac{j}{2} & -\tfrac{j}{2} &  j & - j \\
	- j & -\tfrac{j}{2} &  j &  \tfrac{j}{2} & - j \\
	\tfrac{j}{2} &  j &  \tfrac{j}{2} & - j & - j \\
	-\tfrac{j}{2} & - j & - j & - j & -\tfrac{j}{2}
\end{bsmallmatrix*}
\right)
.
\end{align*}
Matrix ${\mathbf{A}}_{2}$
requires
20~real additions
and $\mathbf{C}_2$
needs
90~real additions and 40~bit-shifting operations.

While
the direct implementation of the $\mathbf{{T}}^*_{31}$
requires
380 additions and 160 bit-shifting operations,
the factorization
presented
in~\eqref{fast11}
needs
130 real additions and 40 bit-shifting operations.
The scaling
to
${\mathbf{F}}^*_{11}$
requires
20 extra multiplications.
In terms of
${\mathbf{F}}^{\prime}_{11}$,
it
needs
40 additions and 40 bit-shifting operations
instead of
20 multiplications of ${\mathbf{F}}^*_{11}$.

\subsection{31-point Approximation Fast Algorithm}\label{31fact}

The low-complexity matrix $\mathbf{T}^*_{31}$
can be expressed as:
\begin{align}
	\label{fast31}
	{\mathbf{T}}^*_{31}=\mathbf{A}_3^\top \cdot \mathbf{C}_3 \cdot \mathbf{A}_3
	,
\end{align}
where
\begin{align*}
\mathbf{A}_3 = \operatorname{diag}(1, \mathbf{B}_{30}).
\end{align*}
The matrix $\mathbf{C}_3$
is
a block matrix given by
\begin{align*}
\mathbf{C}_1 = \operatorname{diag}(\mathbf{E}_1, \mathbf{E}_2)
,
\end{align*}
where
\begin{align*}
	{\mathbf{E}}_{1} =
		\begin{bsmallmatrix*}[r]\\
			1&1&1&1&1&1&1&1&1&1&1&1&1&1&1&1\\
			1&1&1&1&1&\frac{1}{2}&\frac{1}{2}&&&-\frac{1}{2}&-\frac{1}{2}&-\frac{1}{2}&-1&-1&-1&-1\\
			1&1&1&\frac{1}{2}&&-\frac{1}{2}&-1&-1&-1&-1&-\frac{1}{2}&-\frac{1}{2}&&\frac{1}{2}&1&1\\
			1&1&\frac{1}{2}&-\frac{1}{2}&-1&-1&-1&-\frac{1}{2}&&1&1&1&\frac{1}{2}&&-\frac{1}{2}&-1\\
			1&1&&-1&-1&-\frac{1}{2}&&1&1&\frac{1}{2}&-\frac{1}{2}&-1&-1&-\frac{1}{2}&\frac{1}{2}&1\\
			1&\frac{1}{2}&-\frac{1}{2}&-1&-\frac{1}{2}&\frac{1}{2}&1&1&-\frac{1}{2}&-1&-1&&1&1&&-1\\
			1&\frac{1}{2}&-1&-1&&1&\frac{1}{2}&-\frac{1}{2}&-1&&1&1&-\frac{1}{2}&-1&-\frac{1}{2}&1\\
			1&&-1&-\frac{1}{2}&1&1&-\frac{1}{2}&-1&\frac{1}{2}&1&&-1&-\frac{1}{2}&1&\frac{1}{2}&-1\\
			1&&-1&&1&-\frac{1}{2}&-1&\frac{1}{2}&1&-\frac{1}{2}&-1&\frac{1}{2}&1&-\frac{1}{2}&-1&1\\
			1&-\frac{1}{2}&-1&1&\frac{1}{2}&-1&&1&-\frac{1}{2}&-1&1&\frac{1}{2}&-1&&1&-\frac{1}{2}\\
			1&-\frac{1}{2}&-\frac{1}{2}&1&-\frac{1}{2}&-1&1&&-1&1&&-1&1&\frac{1}{2}&-1&\frac{1}{2}\\
			1&-\frac{1}{2}&-\frac{1}{2}&1&-1&&1&-1&\frac{1}{2}&\frac{1}{2}&-1&1&&-1&1&-\frac{1}{2}\\
			1&-1&&\frac{1}{2}&-1&1&-\frac{1}{2}&-\frac{1}{2}&1&-1&1&&-\frac{1}{2}&1&-1&\frac{1}{2}\\
			1&-1&\frac{1}{2}&&-\frac{1}{2}&1&-1&1&-\frac{1}{2}&&\frac{1}{2}&-1&1&-1&1&-\frac{1}{2}\\
			1&-1&1&-\frac{1}{2}&\frac{1}{2}&&-\frac{1}{2}&\frac{1}{2}&-1&1&-1&1&-1&1&-\frac{1}{2}&\\
			1&-1&1&-1&1&-1&1&-1&1&-\frac{1}{2}&\frac{1}{2}&-\frac{1}{2}&\frac{1}{2}&-\frac{1}{2}&&\\
		\end{bsmallmatrix*}
	,
\end{align*}

\begin{align*}
	{\mathbf{E}}_{2} = j \cdot
		\begin{bsmallmatrix*}[r]\\
			-1&1&-1&1&-1&1&-1&1&-\frac{1}{2}&\frac{1}{2}&-\frac{1}{2}&\frac{1}{2}&-\frac{1}{2}&&\\
			1&-1&1&-\frac{1}{2}&&&-\frac{1}{2}&\frac{1}{2}&-1&1&-1&1&-1&\frac{1}{2}&-\frac{1}{2}\\
			-1&1&-\frac{1}{2}&&\frac{1}{2}&-1&1&-1&\frac{1}{2}&&-\frac{1}{2}&1&-1&1&-\frac{1}{2}\\
			1&-\frac{1}{2}&&1&-1&1&&-\frac{1}{2}&1&-1&\frac{1}{2}&\frac{1}{2}&-1&1&-\frac{1}{2}\\
			-1&&\frac{1}{2}&-1&\frac{1}{2}&\frac{1}{2}&-1&1&&-1&1&-\frac{1}{2}&-\frac{1}{2}&1&-1\\
			1&&-1&1&\frac{1}{2}&-1&\frac{1}{2}&\frac{1}{2}&-1&\frac{1}{2}&\frac{1}{2}&-1&&1&-1\\
			-1&-\frac{1}{2}&1&&-1&\frac{1}{2}&\frac{1}{2}&-1&&1&-\frac{1}{2}&-1&1&\frac{1}{2}&-1\\
			1&\frac{1}{2}&-1&-\frac{1}{2}&1&\frac{1}{2}&-1&-\frac{1}{2}&1&\frac{1}{2}&-1&&1&&-1\\
			-\frac{1}{2}&-1&\frac{1}{2}&1&&-1&&1&\frac{1}{2}&-1&-1&\frac{1}{2}&1&-\frac{1}{2}&-1\\
			\frac{1}{2}&1&&-1&-1&\frac{1}{2}&1&\frac{1}{2}&-1&-1&&1&\frac{1}{2}&-\frac{1}{2}&-1\\
			-\frac{1}{2}&-1&-\frac{1}{2}&\frac{1}{2}&1&\frac{1}{2}&-\frac{1}{2}&-1&-1&&1&1&&-1&-1\\
			\frac{1}{2}&1&1&\frac{1}{2}&-\frac{1}{2}&-1&-1&&\frac{1}{2}&1&1&&-\frac{1}{2}&-1&-1\\
			-\frac{1}{2}&-1&-1&-1&-\frac{1}{2}&&1&1&1&\frac{1}{2}&&-\frac{1}{2}&-1&-1&-\frac{1}{2}\\
			&\frac{1}{2}&1&1&1&1&\frac{1}{2}&&-\frac{1}{2}&-\frac{1}{2}&-1&-1&-1&-1&-\frac{1}{2}\\
			&-\frac{1}{2}&-\frac{1}{2}&-\frac{1}{2}&-1&-1&-1&-1&-1&-1&-1&-1&-\frac{1}{2}&-\frac{1}{2}&\\
		\end{bsmallmatrix*}
	.
\end{align*}
The matrix ${\mathbf{A}}_{3}$
requires
only 60 real additions,
while
$\mathbf{C}_3$
needs 780 real additions and 300 bit-shifting operations.

The direct implementation of the $\mathbf{{T}}^*_{31}$
requires
3180 additions and 1200 bit-shifting operations.
After applying the factorization in~\eqref{fast31},
the number of arithmetic operations
is
reduced to
900 real additions and 300 bit-shifting operations.
The scaling to ${\mathbf{F}}^*_{31}$
requires 60 multiplications,
whereas ${\mathbf{F}}^{\prime}_{31}$
needs
120 additions and 120 bit-shifting operations.

\subsection{1023-point {DFT} Approximation}\label{1023-dfttt}

The PFA computation of the 1023-point DFT,
as defined in Section~\ref{S:PFA},
requires
$(33\times 2700) + (93\times 300) + (341\times 12)= 121092$ real multiplications,
and $(33\times 4560) + (93\times 520) + (341\times 24)=207024$ real additions
using
the 3-, 11-, and 31-point DFTs
directly from
their definitions.
However,
if the exact 1023-point DFT
is
computed by
the sparse matrix factorizations detailed in Appendix~\ref{app:exactfast},
then
$(33\times 900) +  (93\times 100)+ (341\times 2)= 39682$ real multiplications,
$(33\times 1020) + (93\times 140) + (341\times 12)=50772$ real additions,
and
$(341\times 2)=682$ bit-shifting operations
are
needed.

The unscaled algorithm
to compute
the 1023-point {DFT} approximation,
referred to
as ${\hat{\mathbf{T}}}^*_{1023}$,
has
null complexity of multiplications.
To calculate
${\hat{\mathbf{T}}}^*_{1023}$,
the matrix ${\mathbf{T}}^*_{31}$
is
called
33 times contributing with $900\times33=29700$ real additions
and $300\times33=9900$ bit-shifting operations.
On the other hand,
${\mathbf{T}}^*_{11}$
is
called
93 times
contributing
with $130\times93=12090$ real additions
and $40\times93=3720$ bit-shifting operations
and ${\mathbf{T}}^*_{3}$
is
called 341 times
which
corresponds to
$12\times341=4092$ real additions and $2\times341=682$ bit-shifting operations.
Then,
the resulting arithmetic costs of ${\hat{\mathbf{T}}}^*_{1023}$
are
45882 real additions and 14302 bit-shifting operations.
To compute
the scaled 1023-point {DFT} approximation
with
the exact $\mathbf{S}$,
called
${\hat{\mathbf{F}}}^*_{1023}$,
more 2044 multiplications are necessary.
However,
if
$\mathbf{S}$
is
approximated following Table~\ref{CSD},
instead of
multiplications,
4088 additions and 4088 bit-shifting operations
are
needed
to achieve ${\hat{\mathbf{F}}}^{\prime}_{1023}$.

\subsection{Hybrid 1023-point {DFT} Approximation}

Applying
the hybrid approach detailed in Section~\ref{S:hybrids}
to
the 1023-point DFT approximation
defined in~\eqref{1023dft} and~\eqref{1023dft-2},
we
obtained
12~distinct approximations.
The 1023-point DFT approximations
consist
of four elements:
a diagonal matrix $\hat{\mathbf{S}}$,
a 3-, an 11-, and a 31-point transformation.
In the hybrid approximations,
these four elements
are
alternated
between exact and approximate form.
Table~\ref{hybrids}
helps
to understand
the approximations
by providing
the combinations of the these four elements.
The computation
of the exact DFT
was performed according
to the algorithms
detailed in the Appendix~\ref{app:exactfast}.

The arithmetic complexity of the 1023-point DFT approximations is summarized in Table~\ref{complexity2} and can be obtained using the following equation:
\begin{align}
\begin{split}
\mathcal{A}(\text{1023-point DFT})
=
\mathcal{A}(\hat{\mathbf{S}})
&
+ 33 \cdot \mathcal{A}(\text{31-point DFT})
\\
&
+ 93 \cdot \mathcal{A}(\text{11-point DFT})
\\
&
+ 341 \cdot \mathcal{A}(\text{3-point DFT})
\end{split}
,
\end{align}
where \( \mathcal{A}(\cdot) \) represents the arithmetic complexity of the argument, including operations such as multiplications, additions, and bit-shifting.

\begin{table}
	\caption{Hybrid Approximations for the 1023-point DFT}
	\centering
	\begin{tabular}{ccccc}
		\toprule
		&&\multicolumn{3}{c}{Employed transformation}\\\cmidrule(rrrrr){3-5}
		Approximation&$\hat{\mathbf{S}}$&3-point&11-point&31-point\\\bottomrule
		${\hat{\mathbf{F}}}^{*}_{1023,\text{I}}$&Exact&$\mathbf{T}^{*}_3$&$\mathbf{F}_{11}$&$\mathbf{F}_{31}$\\
		${\hat{\mathbf{F}}}^{\prime}_{1023,\text{I}}$&CSD approx.&$\mathbf{T}^{*}_3$&$\mathbf{F}_{11}$&$\mathbf{F}_{31}$\\

		${\hat{\mathbf{F}}}^{*}_{1023,\text{II}}$&Exact&$\mathbf{F}_3$&$\mathbf{T}^{*}_{11}$&$\mathbf{F}_{31}$\\
		${\hat{\mathbf{F}}}^{\prime}_{1023,\text{II}}$&CSD approx.&$\mathbf{F}_3$&$\mathbf{T}^{*}_{11}$&$\mathbf{F}_{31}$\\

		${\hat{\mathbf{F}}}^{*}_{1023,\text{III}}$&Exact&$\mathbf{T}^{*}_3$&$\mathbf{T}^{*}_{11}$&$\mathbf{F}_{31}$\\
		${\hat{\mathbf{F}}}^{\prime}_{1023,\text{III}}$&CSD approx.&$\mathbf{T}^{*}_3$&$\mathbf{T}^{*}_{11}$&$\mathbf{F}_{31}$\\

		${\hat{\mathbf{F}}}^{*}_{1023,\text{IV}}$&Exact&$\mathbf{F}_3$&$\mathbf{F}_{11}$&$\mathbf{T}^{*}_{31}$\\
		${\hat{\mathbf{F}}}^{\prime}_{1023,\text{IV}}$&CSD approx.&$\mathbf{F}_3$&$\mathbf{F}_{11}$&$\mathbf{T}^{*}_{31}$\\

		${\hat{\mathbf{F}}}^{*}_{1023,\text{V}}$&Exact&$\mathbf{T}^{*}_3$&$\mathbf{F}_{11}$&$\mathbf{T}^{*}_{31}$\\
		${\hat{\mathbf{F}}}^{\prime}_{1023,\text{V}}$&CSD approx.&$\mathbf{T}^{*}_3$&$\mathbf{F}_{11}$&$\mathbf{T}^{*}_{31}$\\

		${\hat{\mathbf{F}}}^{*}_{1023,\text{VI}}$&Exact&$\mathbf{F}_3$&$\mathbf{T}^{*}_{11}$&$\mathbf{T}^{*}_{31}$\\
		${\hat{\mathbf{F}}}^{\prime}_{1023,\text{VI}}$&CSD approx.&$\mathbf{F}_3$&$\mathbf{T}^{*}_{11}$&$\mathbf{T}^{*}_{31}$\\\bottomrule
	\end{tabular}
	\label{hybrids}
\end{table}

\section{Comparison and Discussion}\label{companddisc}

In this section,
we assess and compare the proposed methods
with
competing methods.
The comparisons
encompass:
arithmetic complexity,
error analysis,
and
frequency response.

\subsection{Arithmetic Complexity}

\subsubsection{Complexity Measurements}

Table~\ref{complexity1}
shows
the arithmetic complexity of
the
proposed
ground transformation matrices
compared
with:
(i)~their respective exact counterparts,
(ii)~the ground transformation used in~\cite{madanayake2020fast} denoted by~$\hat{\mathbf{F}}_{32}$,
and
(iii)~the exact 32-point DFT calculated by
the fully optimized Cooley-Tukey Radix-2~\cite{blahut2010fast} denoted by~${\mathbf{F}}_{32}$.

	\begin{table}
		\caption{Arithmetic Complexity of the Approximate Ground Transforms
and Comparison}
		\label{complexity1}
		\centering
		\begin{tabular}{@{}l@{\hskip 0.05in}l@{\hskip 0in}c@{\hskip 0.05in}c@{\hskip 0.05in}c@{\hskip 0in}}
			\toprule
			$N$&Transform & Real Mult. & Real Add. & Bit-shifting \\\midrule
			\multirow{5}{*}{3}
			& ${\mathbf{F}}_{3}$ (by definition~\cite{oppenheim1999discrete}) & 12 & 24 & 0 \\
			& ${\mathbf{F}}_{3}$ (by the proposed FFT) & 2 & 12 & 2 \\
			& ${\hat{\mathbf{T}}}^*_{3}$ & 0 & 12 & 2 \\
			& ${\hat{\mathbf{F}}}^*_{3}$ & 4 & 12 & 2 \\
			& ${\hat{\mathbf{F}}}^{\prime}_{3}$ & 0 & 20 & 10 \\\midrule
			\multirow{4}{*}{11}
			& ${\mathbf{F}}_{11}$(by definition~\cite{oppenheim1999discrete}) & 300 & 520 & 0 \\
			& ${\mathbf{F}}_{11}$(by the proposed FFT) & 100 & 140 & 0 \\
			& ${\hat{\mathbf{T}}}^*_{11}$ & 0 & 130 & 40 \\
			& ${\hat{\mathbf{F}}}^*_{11}$ & 20 & 130 & 40 \\
			& ${\hat{\mathbf{F}}}^{\prime}_{11}$ & 0 & 170 & 80 \\\midrule
			\multirow{4}{*}{31}
			& ${\mathbf{F}}_{31}$(by definition~\cite{oppenheim1999discrete}) & 2700 & 4560 & 0 \\
			& ${\mathbf{F}}_{31}$(by the proposed FFT) & 900 & 1020 & 0 \\
			& ${\hat{\mathbf{T}}}^*_{31}$ & 0 & 900 & 300 \\
			& ${\hat{\mathbf{F}}}^*_{31}$ & 60 & 900 & 300 \\
			& ${\hat{\mathbf{F}}}^{\prime}_{31}$ & 0 & 1020 & 420 \\\midrule
			\multirow{2}{*}{32 }
			& ${\mathbf{F}}_{32}$(Cooley-Tukey Radix-2~\cite{blahut2010fast}) & 88 & 408 & 0 \\
			& ${\hat{\mathbf{F}}}_{32} $ (proposed in~\cite{madanayake2020fast}) & 0 & 348 & 0 \\
			\bottomrule
		\end{tabular}
	\end{table}

In Table~\ref{complexity2},
we
compare
the arithmetic complexity of the proposed
1023-point DFT approximation algorithm
with
(i)~the 1023-point exact DFT computed by the prime factor algorithm
without fast algorithms for the ground transforms;
(ii)~the 1023-point exact DFT computed by the prime factor algorithm
with fast algorithms for the ground transforms;
(iii)~the 1024-point exact DFT computed by definition~(${\mathbf{F}}_{1024}$);
(iv)~the 1024-point exact DFT computed by the fully optimized Cooley-Tukey \mbox{Radix-2}~(${\mathbf{F}}_{1024}$);
and
(iv)~the fast algorithms for the 1024-point DFT approximation%
---%
${\hat{\mathbf{F}}}_{1024}^\text{I}$,
${\hat{\mathbf{F}}}_{1024}^\text{II}$,
and ${\hat{\mathbf{F}}}_{1024}^\text{III}$%
---%
proposed in~\cite{madanayake2020fast}.
Such methods are
the closest comparable algorithms in the literature.
In ${\hat{\mathbf{F}}}_{1024}^\text{I}$,
both row- and column-wise 32-point DFTs
are
replaced by
the multiplierless ${\hat{\mathbf{F}}}_{32}$~\cite{cintra2024approximation}.
In ${\hat{\mathbf{F}}}_{1024}^\text{II}$ and ${\hat{\mathbf{F}}}_{1024}^\text{III}$,
either the column- or row-wise operation
is kept in the exact form.

\begin{table}

\caption{%
Arithmetic Complexity Assessment and Comparison}
\label{complexity2}
\centering

\begin{tabular}{
  @{}l
  @{\hskip 0.05in}l
  @{\hskip 0in}c
  @{\hskip 0.05in}c
  @{\hskip 0.05in}c
  @{\hskip 0in}
  }
	\toprule

		$N$&Transform & Real Mult. & Real Add. & Bit-shifting
    \\

  \midrule

		\multirow{18}{*}{1023}
		& ${{\mathbf{F}}}_{1023}$ (PFA)~\cite{oppenheim1999discrete})
    & 121092 & 207024 & 0
    \\
		& ${{\mathbf{F}}}_{1023}$ (PFA and fast algorithms)
    & 39682 & 50772 & 682
    \\

  \cline{2-5}

		& ${\hat{\mathbf{F}}}^{*}_{1023,\text{I}}$ & 40364 & 50772 & 682 \\
		& ${\hat{\mathbf{F}}}^{\prime}_{1023,\text{I}}$ & 39000 & 53500 & 3410 \\
		& ${\hat{\mathbf{F}}}^{*}_{1023,\text{II}}$ & 32242 & 49842 & 4402 \\
		& ${\hat{\mathbf{F}}}^{\prime}_{1023,\text{II}}$ & 30382 & 53562 & 8122 \\
		& ${\hat{\mathbf{F}}}^{*}_{1023,\text{III}}$ & 11962 & 46812 & 10582 \\
		& ${\hat{\mathbf{F}}}^{\prime}_{1023,\text{III}}$ & 9982 & 50772 & 14542 \\
		& ${\hat{\mathbf{F}}}^{*}_{1023,\text{IV}}$ & 31684 & 49842 & 4402 \\
		& ${\hat{\mathbf{F}}}^{\prime}_{1023,\text{IV}}$ & 29700 & 53810 & 8370 \\
		& ${\hat{\mathbf{F}}}^{*}_{1023,\text{V}}$ & 11324 & 46812 & 10582 \\
		& ${\hat{\mathbf{F}}}^{\prime}_{1023,\text{V}}$ & 9300 & 50860 & 14630 \\
		& ${\hat{\mathbf{F}}}^{*}_{1023,\text{VI}}$ & 2722 & 45882 & 14302 \\
		& ${\hat{\mathbf{F}}}^{\prime}_{1023,\text{VI}}$ & 682 & 49962 & 18382 \\
		& ${\hat{\mathbf{T}}}^*_{1023}$ & 0 & 45882 & 14302 \\
		& ${\hat{\mathbf{F}}}^*_{1023}$ & 2044 & 45882 & 14302 \\
		& ${\hat{\mathbf{F}}}^{\prime}_{1023}$ & \textbf{0} & \textbf{49970} & \textbf{18390} \\

  \midrule

\multirow{5}{*}{1024}
& \textcolor{black}{${\mathbf{F}}_{1024}$ (exact, by definition~\cite{blahut2010fast})}
& \textcolor{black}{3084288} & \textcolor{black}{5159936} & \textcolor{black}{0} \\
& ${\mathbf{F}}_{1024}$ (Cooley-Tukey~\cite{blahut2010fast}) & 10248 & 30728 & 0 \\
& ${\hat{\mathbf{F}}}_{1024}^\text{I}$(proposed in~\cite{madanayake2020fast}) & 2883 & 25155 & 0 \\
& ${\hat{\mathbf{F}}}_{1024}^\text{II}$(proposed in~\cite{madanayake2020fast}) & 5699 & 27075 & 0 \\
		& ${\hat{\mathbf{F}}}_{1024}^\text{III}$(proposed in~\cite{madanayake2020fast}) & 5699 & 27075 & 0
    \\

  \bottomrule

\end{tabular}
\end{table}

\subsubsection{Discussion}

Table~\ref{complexity1}
shows
that
power-of-two ground transformation matrices
benefit more
from factorization
than
prime-length transformations.
However,
when
ground transforms
are
used as
a building block to derive
larger transforms,
a different phenomenon occurs.
Indeed,
the
$N$~complex multiplications due to the twiddle factors
present
in Cooley-Tukey-based approximations
offset the complexity reductions
from its factorization.
In contrast,
the proposed PFA-based
approximations
do not
require
intermediate
multiplications
by twiddle factors.
In particular,
the proposed
approximations
${\hat{\mathbf{T}}}^*_{1023}$
and
${\hat{\mathbf{F}}}^{\prime}_{1023}$
are
entirely multiplierless.

This feature
has
a direct impact on hardware efficiency.
As a reference,
the approximation with the lowest arithmetic complexity
proposed in~\cite{madanayake2020fast}
(${\hat{\mathbf{F}}}_{1024}^\text{I}$)
achieved reductions of up to 48.5\% in chip area, 30\% in critical path delay~(CPD), and 66.0\% in energy consumption compared to the conventional radix-2 Cooley-Tukey FFT implementation.
Given that the proposed approximations ${\hat{\mathbf{T}}}^*_{1023}$ and ${\hat{\mathbf{F}}}^{\prime}_{1023}$
require no multiplications,
it
is
therefore reasonable to expect
that the proposed method
could achieve
at least comparable reductions
in terms of power consumption, chip area, and CPD.

\subsection{Error Analysis}\label{erroranalysis}

Table~\ref{measures1} and~\ref{measures2}
summarize
the proximity measurements
(Section~\ref{S:optimPro})
for the proposed approximations
compared
the approximations proposed in~\cite{madanayake2020fast,cintra2024approximation}.
Although
the proposed approximations
are not
strictly orthogonal,
their deviations from orthogonality
are
extremely low ($\approx 10^{-2}$).
MAPE measurements
are
also smaller for the proposed approximations.
The total error energy~($\epsilon$)
indicates
that the proposed approximations
are
more refined
than the approximations in~\cite{madanayake2020fast,cintra2024approximation}.
The good performance of the proposed ground approximations
is
transferred to
the 1023-point DFT approximations.

The applicability of approximate DFTs to practical scenarios
such as digital beamforming and spectrum sensing has been demonstrated in~\cite{madanayake2020fast,cintra2024approximation}.
In both cases,
the adopted approximations
proved adequate for real-world implementations
despite presenting higher error levels compared to the methods proposed
in this work.
Therefore,
given that
blocklength is not a critical constraint in these contexts
and that the proposed approximations achieve even lower error rates,
it is reasonable to expect that they are equally, if not more, suitable for these classes of applications,
while fully eliminating multiplications.

\begin{table}
	\caption{Error Measurements of the Ground Approximate Transforms}
	\label{measures1}
	\centering
	\begin{tabular}{lcccc}
		\toprule
		$N$ & Transform & $\epsilon$ & $M$ & $\phi \times 10^{3}$ \\\midrule
		\multirow{2}{*}{3}
		& $\hat{\mathbf{F}}^*_{3}$ & 0.0968 & 1.59 & 6.73 \\
		& $\hat{\mathbf{F}}^{\prime}_{3}$ & 0.0975 & 1.60 & 6.77 \\\midrule
		\multirow{2}{*}{11}
		& $\hat{\mathbf{F}}^*_{11}$ & 8.88 & 1.19 & 14.12 \\
		& $\hat{\mathbf{F}}^{\prime}_{11}$ & 8.90 & 1.20 & 14.11 \\\midrule
		\multirow{2}{*}{31}
		& $\hat{\mathbf{F}}^*_{31}$ & 76.60 & 0.45 & 19.83 \\
		& $\hat{\mathbf{F}}^{\prime}_{31}$ & 76.90 & 0.45 & 19.84 \\\midrule
		\multirow{1}{*}{32}
		& $\hat{\mathbf{F}}_{32}$ & 332 & 0.84 & 36.07 \\
		\bottomrule
	\end{tabular}
\end{table}

\begin{table}
	\caption{Error Measurements of the Proposed 1023-point Approximation and Comparison}
	\label{measures2}
	\centering
	\begin{tabular}{llccc}
		\toprule
		$N$ & Transform & $\epsilon \times 10^{-4}$ & $M \times 10^{3}$ & $\phi \times 10^{3}$ \\\midrule
		\multirow{15}{*}{1023}
		& ${\hat{\mathbf{F}}}^{*}_{1023,\text{I}}$ & 1.13 & 4.67 & 6.73 \\
		& ${\hat{\mathbf{F}}}^{\prime}_{1023,\text{I}}$ & 1.13 & 4.69 & 6.77 \\
		& ${\hat{\mathbf{F}}}^{*}_{1023,\text{II}}$ & 7.68 & 12.83 & 14.12 \\
		& ${\hat{\mathbf{F}}}^{\prime}_{1023,\text{II}}$ & 7.70 & 12.86 & 14.11 \\
		& ${\hat{\mathbf{F}}}^{*}_{1023,\text{III}}$ & 8.35 & 13.68 & 19.83 \\
		& ${\hat{\mathbf{F}}}^{\prime}_{1023,\text{III}}$ & 8.38 & 13.70 & 19.84 \\
		& ${\hat{\mathbf{F}}}^{*}_{1023,\text{IV}}$ & 8.80 & 14.12 & 20.76 \\
		& ${\hat{\mathbf{F}}}^{\prime}_{1023,\text{IV}}$ & 8.88 & 14.18 & 20.79 \\
		& ${\hat{\mathbf{F}}}^{*}_{1023,\text{V}}$ & 9.46 & 14.77 & 26.43 \\
		& ${\hat{\mathbf{F}}}^{\prime}_{1023,\text{V}}$ & 9.55 & 14.82 & 26.49 \\
		& ${\hat{\mathbf{F}}}^{*}_{1023,\text{VI}}$ & 15.93 & 18.67 & 33.68 \\
		& ${\hat{\mathbf{F}}}^{\prime}_{1023,\text{VI}}$ & 16.66 & 19.86 & 33.78 \\
		& $\hat{\mathbf{F}}^*_{1023}$ & 17.03 & 19.41 & 40.18 \\
		& $\hat{\mathbf{F}}^{\prime}_{1023}$ & 17.10 & 19.45 & 40.06 \\\midrule
		\multirow{3}{*}{1024}
		& $\hat{\mathbf{F}}_{1024}^\text{I}$ & 93.00 & 44 & 69.42 \\
		& $\hat{\mathbf{F}}_{1024}^\text{II}$ & 34.02 & 25.31 & 36.07 \\
		& $\hat{\mathbf{F}}_{1024}^\text{III}$ & 34.02 & 25.31 & 36.07 \\
		\bottomrule
	\end{tabular}
\end{table}

\subsection{Frequency Response}

Considering
the rows of the linear transform matrix
as a finite impulse response (FIR) filter bank~\cite{oppenheim1999discrete},
it
is
possible to evaluate the approximation performance
according to
the frequency response of the considered filters.
This approach is justified by the fact that,
in linear time-invariant systems,
any input signal can be expressed as a linear combination of shifted impulses~\cite{mitra2001digital}.
In this way,
analyzing the impulse response of each row provides a complete and interpretable characterization of the transformation behavior.

The analysis
is focused
on
$\hat{\mathbf{F }}^\prime_{3}$,
$\hat{\mathbf{F }}^\prime_{11}$,
$\hat{\mathbf{F }}^\prime_{31}$,
and
$\hat{\mathbf{F }}^\prime_{1023}$
because
they
are
fully approximated, scaled, and free of multiplications.

\subsubsection{Overall Assessment}
The
frequency response
error energy~($\epsilon$)
measurements
for
$\hat{\mathbf{F }}^\prime_{3}$,
$\hat{\mathbf{F }}^\prime_{11}$,
and
$\hat{\mathbf{F }}^\prime_{31}$
relative to the exact filter bank
are
given in Table~\ref{bins}.
The lowest measurements
are
in boldface.

Fig.~1(a)–(d)
shows
the error energy for all filters from the 3-, 11-, 31-, and 1023-point approximations.
Each row of the transform matrix is interpreted as a FIR filter
and is represented by a distinct color in the plot.
The vertical axis corresponds to the normalized magnitude response in decibels (dB), computed as:
\begin{align*}
20 \log_{10} \left( \frac{|H(\omega)|}{\max_\omega |H(\omega)|} \right),
\end{align*}
where $ H(\omega) $ is the frequency response of the row under analysis.
This normalization
sets
the main lobe peak at $ 0\,\text{dB} $, allowing for visual inspection of attenuation and spectral leakage across frequencies.
Although the curves appear visually similar to each other,
it is still possible to observe that
the
error is kept
under~\qty{-17}{\dB}
in all cases.
In~\cite{madanayake2020fast},
the approximations
$\hat{\mathbf{F}}_{1024}^\text{I}$,
$\hat{\mathbf{F}}_{1024}^\text{II}$, and
$\hat{\mathbf{F}}_{1024}^\text{III}$
produced
errors
below~\qty{-6.8}{\dB},~\qty{-11.52}{\dB}, and~\qty{-10.61}{\dB},
respectively .

Notice that,
for all proposed approximations (see Table~\ref{bins}),
including those with blocklength 1023 derived from the PFA,
the DC component
remains
unchanged
and
is computed exactly as in the exact DFT.
This
holds
because neither the approximation procedure nor the orthogonalization process modifies the $0$th row of the DFT matrix,
which is solely composed of ones and inherently has low-complexity.
Therefore,
the approximation error is exclusively associated with the non-DC frequency components,
while the DC level remains identical to the one calculated by exact DFT.
In addition,
an analysis
is
presented for the frequency response to a pure cosine input generated by
\begin{equation}
	x[n] = \cos\left(2\pi \cdot \frac{100}{N} \cdot n\right), \quad n = 0, 1, \dots, N-1
	.
\end{equation}

Fig.~\ref{exponential}
presents
the magnitude responses obtained by applying the exact and DFT approximations
to a pure cosine signal,
where the dashed line
represents
the maximum value of the non-dominant frequencies.
In this analysis,
the results
are
shown in linear scale (rather than in dB) to improve visualization.
As shown in Fig.~\ref{exponential}(a),
the approximation $\hat{\mathbf{F}}^{\text{I}}_{1024}$
exhibits
noticeable leakage across several non-dominant frequencies,
along with distortion in one of the main lobes (left side).
In the hybrid approximation $\hat{\mathbf{F}}^{\text{II}}_{1024}$,
in Fig.~\ref{exponential}(b),
the leakage in non-dominant frequencies
is
reduced
but distortion in the left main lobe remains.
In Fig.~\ref{exponential}(c),
the hybrid approximation $\hat{\mathbf{F}}^{\text{III}}_{1024}$
presents
the main lobes close to their exact counterparts,
but
it
presents
leakage in more non-dominant frequencies that $\hat{\mathbf{F}}^{\text{II}}_{1024}$.
In contrast,
the proposed approximation $\hat{\mathbf{F}}'_{1023}$,
shown in Fig.~\ref{exponential}(d),
achieves
a closer match to the exact DFT in the main lobes
and
a lower maximum leakage in non-dominant frequencies.

\subsubsection{Worst-case Scenario}

Fig.~\ref{figure-freq-response-3-11-31}(a)--(c)
displays
the frequency response magnitude plots of the three least-performing filters for the ground
3-, 11-, and 31--point
approximations
compared with
their exact counterparts,
respectively.

Fig.~\ref{figure-freq-response-1023}(a)--(c)
address
the 1023-point approximation
and
shows the frequency response magnitude plot
associated
to matrix rows (filters)
86,
699,
and 854,
respectively.
These are the least performing filters,
presenting
error energy measurements
of
\num{306.08},
\num{287.1}, and
\num{286.29},
respectively;
being
\num{167.15}
the average error energy from all filters.

Even under such
worst-case scenario analysis,
the approximate methods
were able
to preserve
the main and the secondary lobes from the exact DFT.

\begin{table*}
	\caption{Error Energy of 3-, 11-, and 31-point DFT approximations
(least performing are highlighted)}
	\label{bins}
	\centering
		\setlength{\tabcolsep}{1.5pt}
		\begin{tabular}[t]{c cccc cccc ccc @{\hskip 0.1in}c}
			\toprule
\multirow{2}{*}{Method} & \multicolumn{11}{c}{Row}& \multirow{2}{*}{Total}
\\
\cline{2-12}
&
1&2&3&4&5&6&7&8&9&10&11&
\\
\midrule
			$\hat{\mathbf{F}}^\prime_{3}$&
\textbf{0.00}&\textbf{0.08}&\textbf{0.01}&&&&&&&&&0.09\\
			$\hat{\mathbf{F}}^\prime_{11}$&0.00&0.44&1.01&0.93&\textbf{1.09}&\textbf{1.33}&0.46&0.69&0.85&0.77&\textbf{1.34}&8.90\\
\bottomrule
		\end{tabular}

		\begin{tabular}[t]{c cccc cccc cccc cccc @{\hskip 0.1in}c}
			\toprule
\multirow{2}{*}{Method} & \multicolumn{16}{c}{Row}& \multirow{2}{*}{Total}
\\
\cline{2-17}
&1&2&3&4&5&6&7&8&9&10&11&12&13&14&15&16&
\\
\midrule
\multirow{4}{*}{$\hat{\mathbf{F}}^\prime_{31}$}&
0.00 &2.08&2.91&1.56&\textbf{3.66}&1.97&\textbf{3.69}&3.26&0.82&3.36&1.56&3.38&2.54&1.60&2.73&1.91&
\multirow{4}{*}{76.8}
\\
& \multicolumn{16}{c}{Row}&
\\
\cline{2-17}
			&17&18&19&20&21&22&23&24&25&26&27&28&29&30&31&
\\
&
3.20&2.38&3.51&2.57&1.73&3.56&1.77&\textbf{4.29}&1.87&1.45&3.16&1.47&3.55&2.22&3.04&&
\\
\bottomrule
		\end{tabular}

\end{table*}

\begin{figure*}[t]
	\psfrag{XXXXX}[c][c]{\scriptsize Normalized Frequency}
	\psfrag{YYYYY}[c][c]{\scriptsize $\text{Magnitude (dB)}$}
	\psfrag{aaaaa}[c][c]{\scriptsize $-\pi$}
	\psfrag{bbbbb}[c][c]{\scriptsize $-\pi/2$}
	\psfrag{ccccc}[c][c]{\scriptsize $0$}
	\psfrag{ddddd}[c][c]{\scriptsize $\pi/2$}
	\psfrag{eeeee}[c][c]{\scriptsize $\pi$}

	\psfrag{-40}[c][c]{\scriptsize $-40$}
	\psfrag{-30}[c][c]{\scriptsize $-30$}
	\psfrag{-10}[c][c]{\scriptsize $-10$}
	\psfrag{0}[c][c]{\scriptsize $0$}
	\psfrag{-22.86}[c][c]{\scriptsize $-22.86$}
	\psfrag{-17.69}[c][c]{\scriptsize $-17.69$}
	\psfrag{-19.91}[c][c]{\scriptsize $-19.91$}
	\psfrag{-20.9}[c][c]{\scriptsize $-20.9$}

	\centering
	\subfloat[$\hat{\mathbf{F }}^\prime_{3}$]{%
		\includegraphics[width=0.24\linewidth]{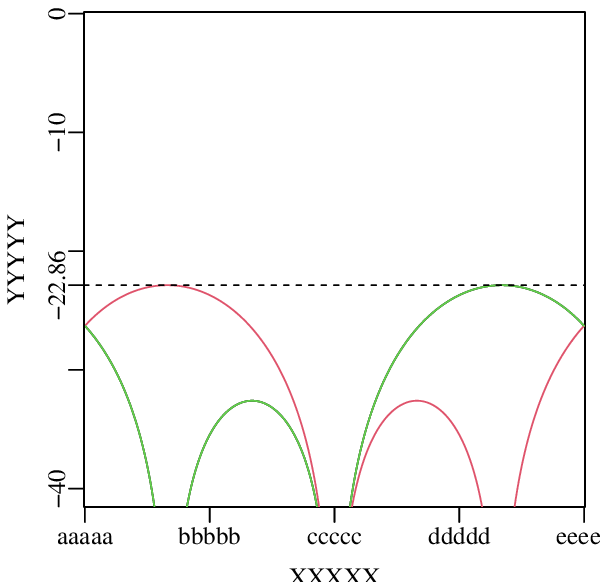}
		\label{e3}
	}\hfil
	\subfloat[$\hat{\mathbf{F }}^\prime_{11}$]{%
		\includegraphics[width=0.24\linewidth]{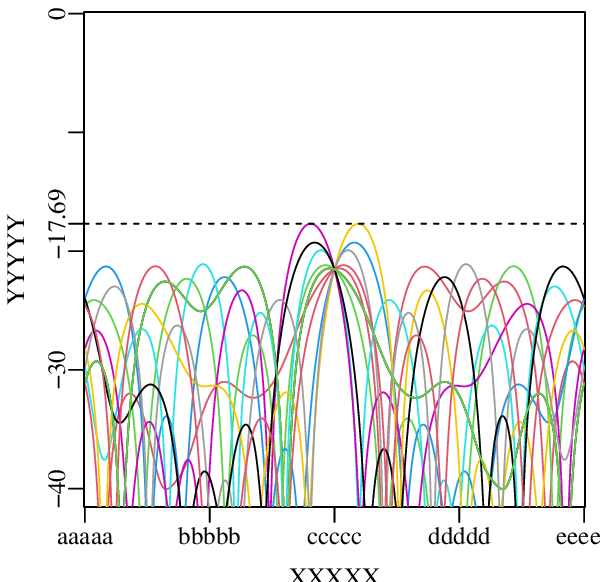}
		\label{e11}
	}\hfil
	\subfloat[$\hat{\mathbf{F }}^\prime_{31}$]{%
		\includegraphics[width=0.24\linewidth]{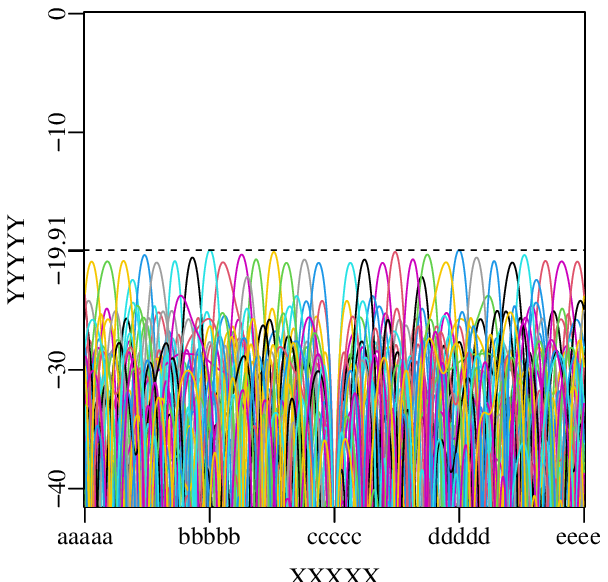}
		\label{e31}
	}\hfil
	\subfloat[$\hat{\mathbf{F }}^\prime_{1023}$]{%
		\includegraphics[width=0.24\linewidth]{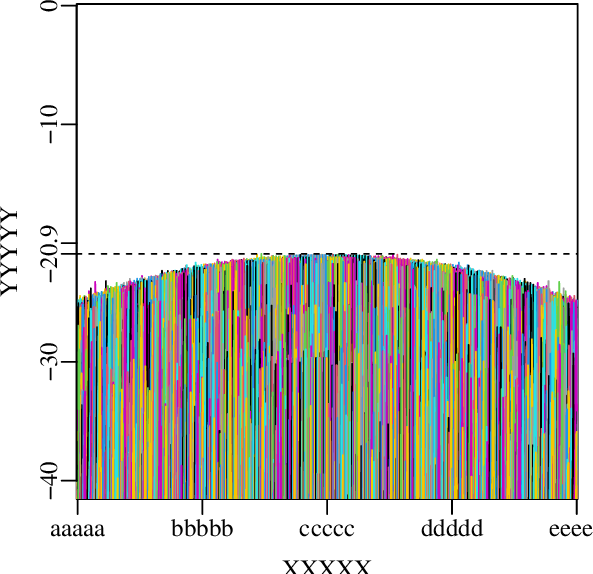}
		\label{e1023}
	}

	\caption{Error plots between the filter bank frequency response magnitude for (a)~3-point approximation, (b)~11-point approximation, (c)~31-point approximation, and (d)~1023-point approximation and their exact counterparts.}
	\label{figure-response-full}
\end{figure*}

\begin{figure}[htbp]
	\psfrag{XXXXX}[c][c]{\footnotesize Normalized Frequency}
	\psfrag{YYYYY}[c][c]{\footnotesize $\text{Magnitude (dB)}$}
	\psfrag{aaaaa}[c][c]{\footnotesize $-\pi$}
	\psfrag{bbbbb}[c][c]{\footnotesize $-\pi/2$}
	\psfrag{ccccc}[c][c]{\footnotesize $0$}
	\psfrag{ddddd}[c][c]{\footnotesize $\pi/2$}
	\psfrag{eeeee}[c][c]{\footnotesize $\pi$}
	\psfrag{RR9}[c][c]{\footnotesize \hspace{-1.1mm} Row 24}
	\psfrag{RR8}[c][c]{\footnotesize \hspace{-1.5mm}Row 7}
	\psfrag{RR7}[c][c]{\footnotesize \hspace{-1.5mm}Row 5}
	\psfrag{RR6}[c][c]{\footnotesize \hspace{-1.1mm} Row 11}
	\psfrag{RR5}[c][c]{\footnotesize \hspace{-1.5mm}Row 6}
	\psfrag{RR4}[c][c]{\footnotesize \hspace{-1.5mm}Row 5}
	\psfrag{RR3}[c][c]{\footnotesize \hspace{-1.5mm}Row 3}
	\psfrag{RR2}[c][c]{\footnotesize \hspace{-1.5mm}Row 2}
	\psfrag{RR1}[c][c]{\footnotesize \hspace{-1.5mm}Row 1}
	\psfrag{RR0}[c][c]{\footnotesize \hspace{-1.5mm}Exact}
	\psfrag{-40}[c][c]{\footnotesize $-40$}
	\psfrag{-30}[c][c]{\footnotesize $-30$}
	\psfrag{-20}[c][c]{\footnotesize $-20$}
	\psfrag{-10}[c][c]{\footnotesize $-10$}
	\psfrag{0}[c][c]{\footnotesize $0$}

	\centering
	\subfloat[$\hat{\mathbf{F }}^\prime_{3}$]{%
		\includegraphics[width=0.4\textwidth]{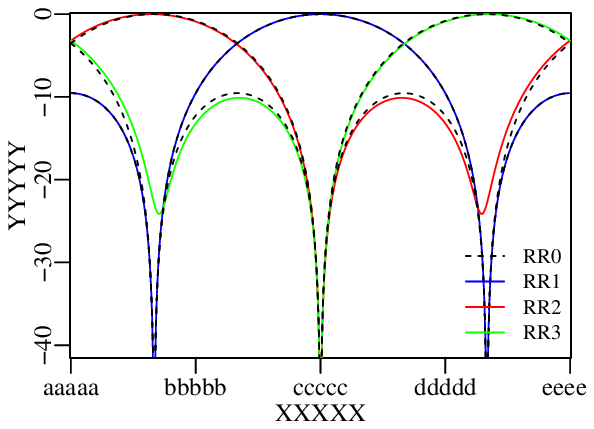}
		\label{w3}
	}
\\
	\subfloat[$\hat{\mathbf{F }}^\prime_{11}$]{%
		\includegraphics[width=0.4\textwidth]{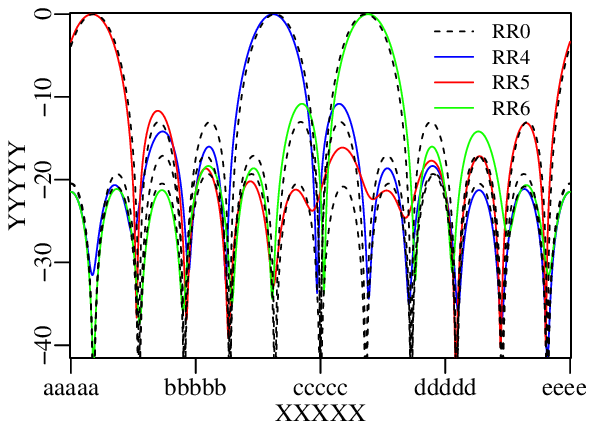}
		\label{w11}
	}
\\
	\subfloat[$\hat{\mathbf{F }}^\prime_{31}$]{%
		\includegraphics[width=0.4\textwidth]{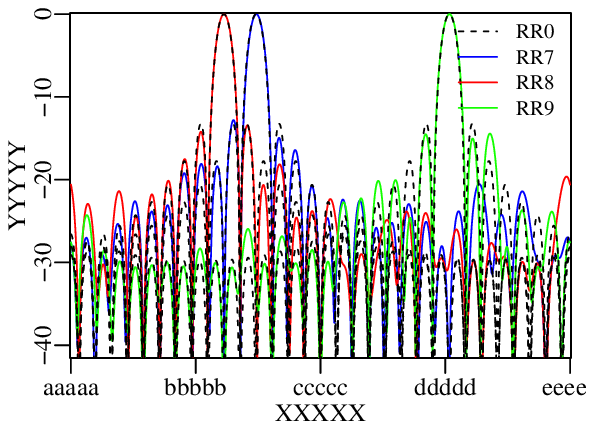}
		\label{w31}
	}

	\caption{Comparison between the magnitude of the filter-bank responses of the approximations and their exact counterparts for the least three performing rows.}
	\label{figure-freq-response-3-11-31}
\end{figure}

\begin{figure*}[htbp]
	\centering

	\subfloat[Row 854]{%
	\psfrag{XXXXXXXXXXXXXXXXXX}{\footnotesize $\text{Normalized Frequency}$}
	\psfrag{YYYYYYYYYYYYYY}{\hspace{0.3cm} \footnotesize $\text{Magnitude (dB)}$}
	\psfrag{EXX}{\footnotesize $\mathbf{F}_{1023} $}
	\psfrag{FFF}{\footnotesize $\hat{\mathbf{F }}^\prime_{1023} $}
	\psfrag{ee}{\footnotesize $\pi/6$}
	\psfrag{jj}{\footnotesize $\pi/4$}
	\psfrag{ll}{\footnotesize $\pi/3$}
	\psfrag{kk}{\footnotesize $5\pi/12$}
	\psfrag{ii}{\footnotesize $\pi/2$}
	\includegraphics[width=0.32\textwidth]{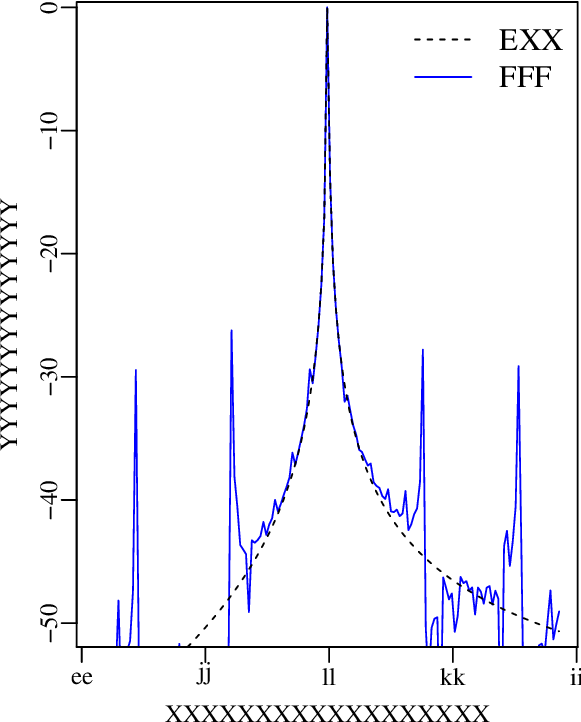}
}
	\hfill
    \subfloat[\footnotesize Row 699]{%
	\psfrag{XXXXXXXXXXXXXXXXXX}{\hspace{0.3cm} \footnotesize $\text{Normalized Frequency}$}
	\psfrag{YYYYYYYYYYYYYY}{\hspace{0.3cm} \footnotesize $\text{Magnitude (dB)}$}
	\psfrag{EXX}{\footnotesize $\mathbf{F}_{1023} $}
	\psfrag{FFF}{\footnotesize $\hat{\mathbf{F }}^\prime_{1023} $}
	    \psfrag{ii}{\footnotesize $\pi/2$}
	\psfrag{mm}{\footnotesize $7\pi/12$}
	\psfrag{nn}{\footnotesize $2\pi/3$}
	\psfrag{pp}{\footnotesize $3\pi/4$}
	\psfrag{ee}{\footnotesize $\pi/6$}
	\psfrag{jj}{\footnotesize $\pi/4$}
	\psfrag{ll}{\footnotesize $\pi/3$}
	\psfrag{kk}{\footnotesize $5\pi/12$}
	\psfrag{ii}{\footnotesize $\pi/2$}
	\includegraphics[width=0.32\textwidth]{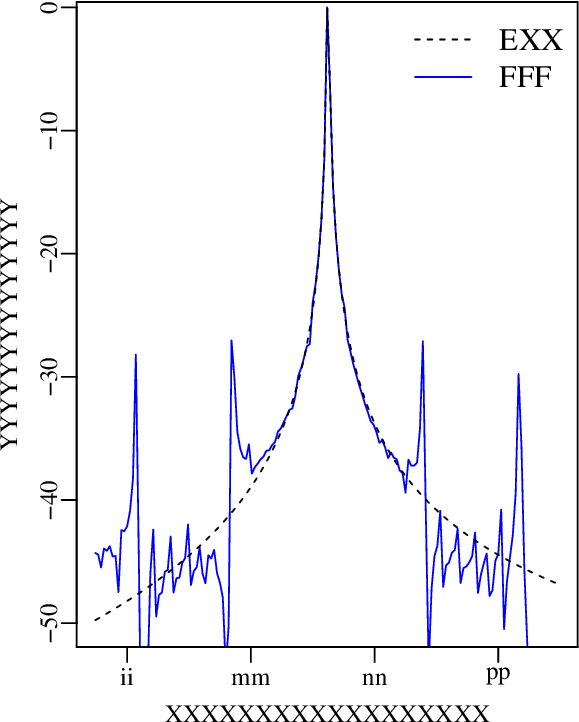}
}
	\hfill
	\subfloat[\footnotesize Row 86]{%
	\psfrag{XXXXXXXXXXXXXXXXXX}{\hspace{0.3cm} \footnotesize $\text{Normalized Frequency}$}
	\psfrag{YYYYYYYYYYYYYY}{\hspace{0.3cm} \footnotesize $\text{Magnitude (dB)}$}
	\psfrag{EXX}{\footnotesize $\mathbf{F}_{1023} $}
	\psfrag{FFF}{\footnotesize $\hat{\mathbf{F }}^\prime_{1023} $}
    \psfrag{aa}{\footnotesize $-\pi/3$}
	\psfrag{dd}{\footnotesize $-\pi/4$}
	\psfrag{cc}{\footnotesize $-\pi/6$}
	\psfrag{bb}{\footnotesize $-\pi/12$}
	\psfrag{hh}{\footnotesize $0$}
		\includegraphics[width=0.32\textwidth]{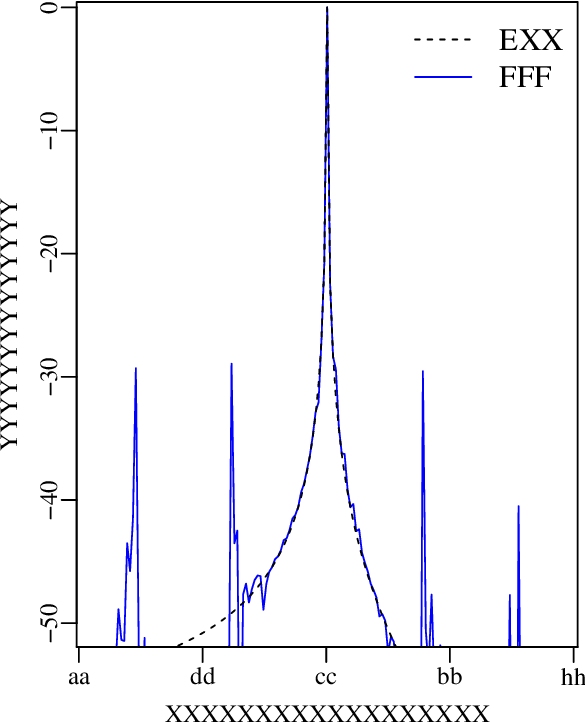}
	}

	\caption{Comparison between the magnitude of the filter-bank responses of the approximation $\hat{\mathbf{F }}^\prime_{1023} $ and the exact {DFT} for the least three performing rows.}
	\label{figure-freq-response-1023}
\end{figure*}

\begin{figure*}[htbp]
	\centering

	\psfrag{XXXXX}[c][c]{\scriptsize DFT index}
	\psfrag{YYYYY}[c][c]{\scriptsize Magnitude}
	\psfrag{AAAA}[l][l]{\scriptsize $\hat{\mathbf{F }}^\prime_{1023} $}
	\psfrag{BBBB}[l][l]{\scriptsize $\mathbf{F}_{1023} $}
	\psfrag{CCCC}[l][l]{\scriptsize $\hat{\mathbf{F }}^{\text{I}}_{1024} $}
	\psfrag{DDDD}[l][l]{\scriptsize $\mathbf{F}_{1024} $}
	\psfrag{EEEE}[l][l]{\scriptsize $\hat{\mathbf{F }}^{\text{II}}_{1024} $}
	\psfrag{FFFF}[l][l]{\scriptsize $\hat{\mathbf{F }}^{\text{III}}_{1024} $}
	\psfrag{-45}[1][1]{\scriptsize $-45$}

	\psfrag{rrr}[1][1]{\scriptsize $0.11$}
	\psfrag{ttt}[1][1]{\scriptsize $0.17$}
	\psfrag{uuuu}[1][1]{\scriptsize $0.17$}
	\psfrag{zzz}[1][1]{\scriptsize $0.09$}
	\psfrag{200}[1][1]{\scriptsize $200$}
	\psfrag{400}[1][1]{\scriptsize $400$}
	\psfrag{600}[1][1]{\scriptsize $600$}
	\psfrag{800}[1][1]{\scriptsize $800$}
	\psfrag{1000}[1][1]{\scriptsize $1000$}
	\psfrag{512}[1][1]{\scriptsize $512$}
	\psfrag{618}[1][1]{\scriptsize $618$}
	\psfrag{-128}[1][1]{\footnotesize $-128$}
	\psfrag{116.5}[1][1]{\footnotesize $116.5$}
	\psfrag{746}[1][1]{\footnotesize $746$}
	\psfrag{96.5}[1][1]{\footnotesize $96.5$}

	\psfrag{0}[1][1]{\footnotesize $0$}
	\psfrag{0.4}[1][1]{\footnotesize $0.4$}
	\psfrag{0.6}[1][1]{\footnotesize $0.6$}
	\psfrag{0.8}[1][1]{\footnotesize $0.8$}
	\psfrag{1}[1][1]{\footnotesize $1$}
	\psfrag{0.09}[1][1]{\footnotesize $0.09$}
	\psfrag{0.17}[1][1]{\footnotesize $0.17$}
	\psfrag{0.11}[1][1]{\footnotesize $0.11$}

	\subfloat[]{
		\includegraphics[width=0.235\linewidth]{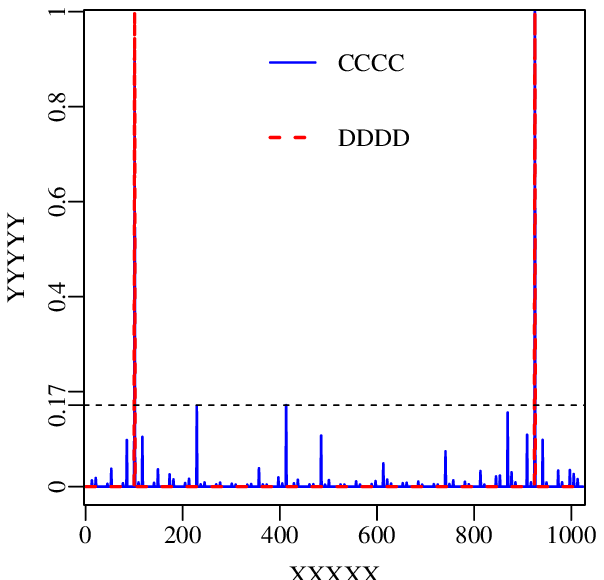}
	}\hfil
	\subfloat[]{
		\includegraphics[width=0.235\linewidth]{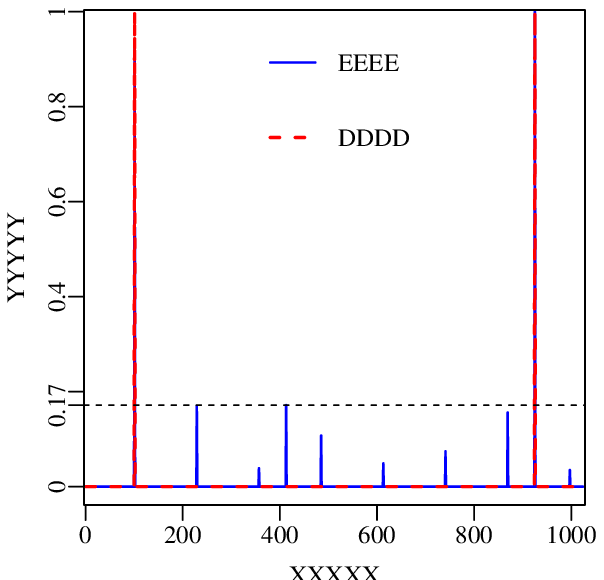}
	}\hfil
	\subfloat[]{
		\includegraphics[width=0.235\linewidth]{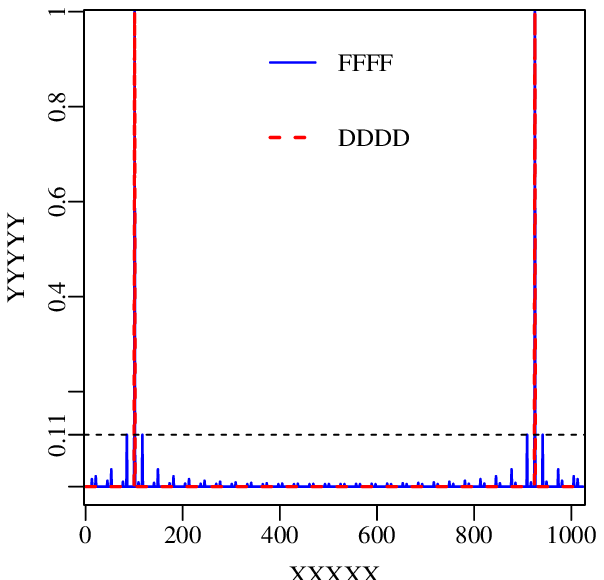}
	}\hfil
	\subfloat[]{
		\includegraphics[width=0.235\linewidth]{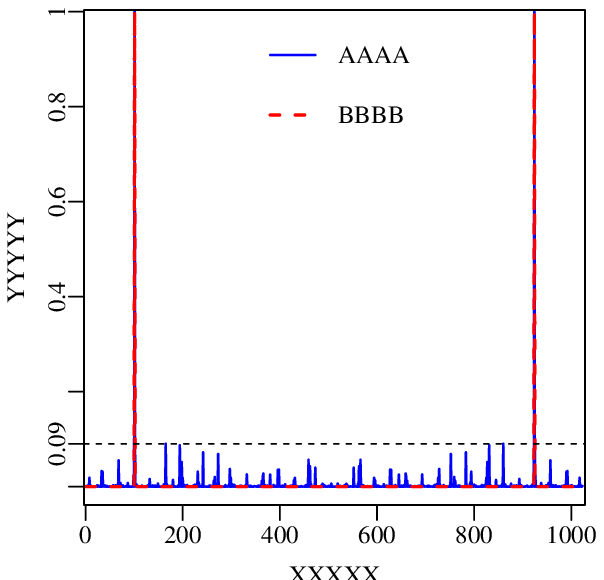}
	}

	\caption{Comparison between the magnitude responses obtained from~\cite{madanayake2020fast} and $\hat{\mathbf{F}}'_{1023}$ with their exact counterparts, applied to a pure cosine signal.}
	\label{exponential}
\end{figure*}

\section{Conclusion}\label{S:conclusion}

In this paper,
we
proposed
a method
to obtain
matrices with null multiplicative complexity
on the prime~factor algorithm.
We
demonstrated
that,
if
the transform length
can
be decomposed into
relatively prime numbers,
then
the entire approximate DFT computation
can
be performed
multiplierlessly.
The absence of twiddle factors
simplifies the design
when compared to Cooley-Tukey-based approximations
and
also
reduces
the error propagation
from the
approximate ground transforms.
We
applied
the proposed method to derive a 1023-point DFT approximation
along with
a collection of approximations,
trading-off
accuracy for computational cost and vice-versa.
The proposed method
outperformed
the comparable methods
in the literature according to popular
figures of merit.

The main contributions of this paper are summarized as follows:
\begin{itemize}
	\item A generic method to obtain multiplierless DFT approximations using the prime factor algorithm;
	\item Three novel approximations for the 3-point DFT;
	\item Three novel approximations for the 11-point DFT;
	\item Three novel approximations for the 31-point DFT;
	\item A procedure to construct large multiplierless DFT approximations when the transform length can be decomposed into relatively prime factors;
	\item Fifteen approximations for the 1023-point DFT, derived using the proposed method.
\end{itemize}

For future work,
we
aim to extend
this framework to different blocklengths.
Preliminary results for the $65$-point DFT ($65=5\times13$)
have already corroborated the scalability
and
consistency of the method, preserving its fully multiplierless and low-complexity characteristics.
In particular,
large transform sizes that result from coprime factorizations---such as 2046 ($2\times3\times11\times31$)---which directly benefits from
the results and approximations developed in this work.
Additionally,
practical hardware implementations of the proposed method
is sought to be investigated.
In this context,
we
highlight
memory-efficient strategies such as in-place and in-order computation.
As shown in~\cite{lun1993analysis},
such techniques
are
directly possible in PFA algorithms only when
the transform blocklength prime factors are quadratic residues of each other.
However,
an earlier work by~\cite{burrus1981place}
introduced
a variation of the PFA that enables
fully in-place and in-order execution for arbitrary factorizations.
This approach significantly
reduces
memory usage, eliminates the need for post-processing (unscrambling), and may achieve faster computation than traditional FFT algorithms such as Cooley-Tukey radix-4.

\appendix
\section{Fast Algorithms for the Exact Transforms}
\label{app:exactfast}

This appendix
discusses
the arithmetic complexity
and
provides
fast algorithms
for the exact 3-, 11-, and 31-point DFT.
The resulting arithmetic complexity is shown in Table~\ref{complexity1}.

\subsection{$3$-, $11$-, and $31$-point DFT}\label{bydef}

The direct computation of the DFT
requires
complex multiplications in the order of $\mathcal{O}(N^2)$~\cite{oppenheim1999discrete}.
Therefore,
disregarding the trivial multipliers,
if
we
consider
a complex input,
the direct computation of the 3-, 11-, and 31-point DFT
requires
\num{12}~real multiplications and \num{24}~real additions;
\num{300}~real multiplications and \num{520}~real additions;
and
\num{2700}~real multiplications and \num{4560}~real additions.
respectively.

\subsection{Fast Algorithms}

\subsubsection{3-point DFT}
Matrix $\mathbf{F}_{3}$ can be factorized as:
\begin{align}
	\label{fast3exact}
	{\mathbf{F}}_{3}=\mathbf{A}_1^\top \cdot
\begin{bsmallmatrix*}[r]\\
	1&1&\\
	1&-\frac{1}{2}&\\
	&&\frac{-j\sqrt{3}}{2}\\
\end{bsmallmatrix*}
\cdot \mathbf{A}_1
.
\end{align}

\subsubsection{11-point DFT}
Let $\beta_x=\cos \left(\frac{x\cdot\pi}{N}\right)$
and  $\delta_x=j\cdot\sin\left(\frac{x\cdot\pi}{N}\right)$.
Matrix $\mathbf{F}_{11}$
can be represented as:
\begin{align}
	\label{fast11exact}
	{\mathbf{F}}_{11}=\mathbf{A}_2^\top\cdot
	\begin{bsmallmatrix*}[r]
		1 & 1 & 1 & 1 & 1 & 1 & & & & & \\
		1 & \beta_2 & \beta_4 & \beta_6 & \beta_8 & \beta_{10} & & & & & \\
		1 & \beta_4 & \beta_8 & \beta_{10} & \beta_6 & \beta_2 & & & & & \\
		1 & \beta_6 & \beta_{10} & \beta_4 & \beta_2 & \beta_8 & & & & & \\
		1 & \beta_8 & \beta_6 & \beta_2 & \beta_{10} & \beta_4 & & & & & \\
		1 & \beta_{10} & \beta_2 & \beta_8 & \beta_4 & \beta_6 & & & & & \\
		& & & & & & -\delta_{5} & \delta_4 & -\delta_3 & \delta_2 & -\delta_1 \\
		& & & & & & \delta_{4} & -\delta_{1} & -\delta_2 & \delta_5 & -\delta_{3} \\
		& & & & & & -\delta_3 & -\delta_2 & \delta_4 & \delta_{1} & -\delta_ 5\\
		& & & & & & \delta_2 & \delta_5 & \delta_{1} & -\delta_3 & -\delta_4 \\
		& & & & & & -\delta_{1} & -\delta_{3} & -\delta_5 & -\delta_4 & -\delta_2 \\
	\end{bsmallmatrix*}
\cdot\mathbf{A}_2
.
\end{align}

\subsubsection{31-point DFT}

Matrix $\mathbf{F}_{31}$ can be expressed as:
\begin{align}
	\label{fast31exact}
	{\mathbf{F}}_{31}=\mathbf{A}_3^\top\cdot
\begin{bsmallmatrix*}
		{\mathbf{E}}_{3}&\\
		&{\mathbf{E}}_{4}
\end{bsmallmatrix*}
\cdot\mathbf{A}_3
	,
\end{align}
where
\begin{align*}
	{\mathbf{E}}_{3} =
		\begin{bsmallmatrix*}[r]
			1 & 1 & 1 & 1 & 1 & 1 & 1 & 1 & 1 & 1 & 1 & 1 & 1 & 1 & 1 & 1 \\
			1 & \beta_2 & \beta_4 & \beta_6 & \beta_8 & \beta_{10} & \beta_{12} & \beta_{14} & \beta_{16} & \beta_{18} & \beta_{20} & \beta_{22} & \beta_{24} & \beta_{26} & \beta_{28} & \beta_{30} \\
			1 & \beta_4 & \beta_8 & \beta_{12} & \beta_{16} & \beta_{20} & \beta_{24} & \beta_{28} & \beta_{30} & \beta_{26} & \beta_{22} & \beta_{18} & \beta_{14} & \beta_{10} & \beta_6 & \beta_2 \\
			1 & \beta_6 & \beta_{12} & \beta_{18} & \beta_{24} & \beta_{30} & \beta_{26} & \beta_{20} & \beta_{14} & \beta_8 & \beta_2 & \beta_4 & \beta_{10} & \beta_{16} & \beta_{22} & \beta_{28} \\
			1 & \beta_8 & \beta_{16} & \beta_{24} & \beta_{30} & \beta_{22} & \beta_{14} & \beta_6 & \beta_2 & \beta_{10} & \beta_{18} & \beta_{26} & \beta_{28} & \beta_{20} & \beta_{12} & \beta_4 \\
			1 & \beta_{10} & \beta_{20} & \beta_{30} & \beta_{22} & \beta_{12} & \beta_2 & \beta_8 & \beta_{18} & \beta_{28} & \beta_{24} & \beta_{14} & \beta_4 & \beta_6 & \beta_{16} & \beta_{26} \\
			1 & \beta_{12} & \beta_{24} & \beta_{26} & \beta_{14} & \beta_2 & \beta_{10} & \beta_{22} & \beta_{28} & \beta_{16} & \beta_4 & \beta_8 & \beta_{20} & \beta_{30} & \beta_{18} & \beta_6 \\
			1 & \beta_{14} & \beta_{28} & \beta_{20} & \beta_6 & \beta_8 & \beta_{22} & \beta_{26} & \beta_{12} & \beta_2 & \beta_{16} & \beta_{30} & \beta_{18} & \beta_4 & \beta_{10} & \beta_{24} \\
			1 & \beta_{16} & \beta_{30} & \beta_{14} & \beta_2 & \beta_{18} & \beta_{28} & \beta_{12} & \beta_4 & \beta_{20} & \beta_{26} & \beta_{10} & \beta_6 & \beta_{22} & \beta_{24} & \beta_8 \\
			1 & \beta_{18} & \beta_{26} & \beta_8 & \beta_{10} & \beta_{28} & \beta_{16} & \beta_2 & \beta_{20} & \beta_{24} & \beta_6 & \beta_{12} & \beta_{30} & \beta_{14} & \beta_4 & \beta_{22} \\
			1 & \beta_{20} & \beta_{22} & \beta_2 & \beta_{18} & \beta_{24} & \beta_4 & \beta_{16} & \beta_{26} & \beta_6 & \beta_{14} & \beta_{28} & \beta_8 & \beta_{12} & \beta_{30} & \beta_{10} \\
			1 & \beta_{22} & \beta_{18} & \beta_4 & \beta_{26} & \beta_{14} & \beta_8 & \beta_{30} & \beta_{10} & \beta_{12} & \beta_{28} & \beta_6 & \beta_{16} & \beta_{24} & \beta_2 & \beta_{20} \\
			1 & \beta_{24} & \beta_{14} & \beta_{10} & \beta_{28} & \beta_4 & \beta_{20} & \beta_{18} & \beta_6 & \beta_{30} & \beta_8 & \beta_{16} & \beta_{22} & \beta_2 & \beta_{26} & \beta_{12} \\
			1 & \beta_{26} & \beta_{10} & \beta_{16} & \beta_{20} & \beta_6 & \beta_{30} & \beta_4 & \beta_{22} & \beta_{14} & \beta_{12} & \beta_{24} & \beta_2 & \beta_{28} & \beta_8 & \beta_{18} \\
			1 & \beta_{28} & \beta_6 & \beta_{22} & \beta_{12} & \beta_{16} & \beta_{18} & \beta_{10} & \beta_{24} & \beta_4 & \beta_{30} & \beta_2 & \beta_{26} & \beta_8 & \beta_{20} & \beta_{14} \\
			1 & \beta_{30} & \beta_2 & \beta_{28} & \beta_4 & \beta_{26} & \beta_6 & \beta_{24} & \beta_8 & \beta_{22} & \beta_{10} & \beta_{20} & \beta_{12} & \beta_{18} & \beta_{14} & \beta_{16} \\
		\end{bsmallmatrix*}
	,
\end{align*}
\begin{align*}
	{\mathbf{E}}_{4}=
		\begin{bsmallmatrix*}[r]
			-\delta_{15}&  \delta_{14} & -\delta_{13} &  \delta_{12} & -\delta_{11} &  \delta_{10} & -\delta_{9} &  \delta_{8} & -\delta_{7} &  \delta_{6} & -\delta_{5} &  \delta_{4} & -\delta_{3} &  \delta_{2} & -\delta_{1} \\
			\delta_{14} & -\delta_{11} &  \delta_{8} & -\delta_{5} &  \delta_{2} &  \delta_{1} & -\delta_{4} &  \delta_{7} & -\delta_{10} &  \delta_{13} & -\delta_{15}&  \delta_{12} & -\delta_{9} &  \delta_{6} & -\delta_{3} \\
			-\delta_{13} &  \delta_{8} & -\delta_{3} & -\delta_{2} &  \delta_{7} & -\delta_{12} &  \delta_{14} & -\delta_{9} &  \delta_{4} &  \delta_{1} & -\delta_{6} &  \delta_{11} & -\delta_{15}&  \delta_{10} & -\delta_{5} \\
			\delta_{12} & -\delta_{5} & -\delta_{2} &  \delta_{9} & -\delta_{15}&  \delta_{8} & -\delta_{1} & -\delta_{6} &  \delta_{13} & -\delta_{11} &  \delta_{4} &  \delta_{3} & -\delta_{10} &  \delta_{14} & -\delta_{7} \\
			-\delta_{11} &  \delta_{2} &  \delta_{7} & -\delta_{15}&  \delta_{6} &  \delta_{3} & -\delta_{12} &  \delta_{10} & -\delta_{1} & -\delta_{8} &  \delta_{14} & -\delta_{5} & -\delta_{4} &  \delta_{13} & -\delta_{9} \\
			\delta_{10} &  \delta_{1} & -\delta_{12} &  \delta_{8} &  \delta_{3} & -\delta_{14} &  \delta_{6} &  \delta_{5} & -\delta_{15}&  \delta_{4} &  \delta_{7} & -\delta_{13} &  \delta_{2} &  \delta_{9} & -\delta_{11} \\
			-\delta_{9} & -\delta_{4} &  \delta_{14} & -\delta_{1} & -\delta_{12} &  \delta_{6} &  \delta_{7} & -\delta_{11} & -\delta_{2} &  \delta_{15}& -\delta_{3} & -\delta_{10} &  \delta_{8} &  \delta_{5} & -\delta_{13} \\
			\delta_{8} &  \delta_{7} & -\delta_{9} & -\delta_{6} &  \delta_{10} &  \delta_{5} & -\delta_{11} & -\delta_{4} &  \delta_{12} &  \delta_{3} & -\delta_{13} & -\delta_{2} &  \delta_{14} &  \delta_{1} & -\delta_{15}\\
			-\delta_{7} & -\delta_{10} &  \delta_{4} &  \delta_{13} & -\delta_{1} & -\delta_{15}& -\delta_{2} &  \delta_{12} &  \delta_{5} & -\delta_{9} & -\delta_{8} &  \delta_{6} &  \delta_{11} & -\delta_{3} & -\delta_{14} \\
			\delta_{6} &  \delta_{13} &  \delta_{1} & -\delta_{11} & -\delta_{8} &  \delta_{4} &  \delta_{15}&  \delta_{3} & -\delta_{9} & -\delta_{10} &  \delta_{2} &  \delta_{14} &  \delta_{5} & -\delta_{7} & -\delta_{12} \\
			-\delta_{5} & -\delta_{15}& -\delta_{6} &  \delta_{4} &  \delta_{14} &  \delta_{7} & -\delta_{3} & -\delta_{13} & -\delta_{8} &  \delta_{2} &  \delta_{12} &  \delta_{9} & -\delta_{1} & -\delta_{11} & -\delta_{10} \\
			\delta_{4} &  \delta_{12} &  \delta_{11} &  \delta_{3} & -\delta_{5} & -\delta_{13} & -\delta_{10} & -\delta_{2} &  \delta_{6} &  \delta_{14} &  \delta_{9} &  \delta_{1} & -\delta_{7} & -\delta_{15}& -\delta_{8} \\
			-\delta_{3} & -\delta_{9} & -\delta_{15}& -\delta_{10} & -\delta_{4} &  \delta_{2} &  \delta_{8} &  \delta_{14} &  \delta_{11} &  \delta_{5} & -\delta_{1} & -\delta_{7} & -\delta_{13} & -\delta_{12} & -\delta_{6} \\
			\delta_{2} &  \delta_{6} &  \delta_{10} &  \delta_{14} &  \delta_{13} &  \delta_{9} &  \delta_{5} &  \delta_{1} & -\delta_{3} & -\delta_{7} & -\delta_{11} & -\delta_{15}& -\delta_{12} & -\delta_{8} & -\delta_{4} \\
			-\delta_{1} & -\delta_{3} & -\delta_{5} & -\delta_{7} & -\delta_{9} & -\delta_{11} & -\delta_{13} & -\delta_{15}& -\delta_{14} & -\delta_{12} & -\delta_{10} & -\delta_{8} & -\delta_{6} & -\delta_{4} & -\delta_{2} \\
		\end{bsmallmatrix*}
.
\end{align*}

\onecolumn

{\small
\singlespacing
\bibliographystyle{siam}
\bibliography{ref}
}

\end{document}